\documentclass{article}
\usepackage[utf8]{inputenc}
\usepackage[margin = 1in]{geometry}

\usepackage{amsmath}
\usepackage{bm}
\usepackage{graphicx}
\usepackage{xcolor}
\usepackage{hyperref}
\usepackage{dsfont}
\def\b{\ensuremath\mathbf}

\usepackage[toc,page]{appendix}

\usepackage[round]{natbib}

\DeclareMathOperator*{\argmin}{argmin}
\DeclareMathOperator{\sign}{sign}

\providecommand{\keywords}[1]
{
  \small	
  \textbf{\textit{Keywords---}} #1
}
      
\title{Transitions between peace and systemic war as bifurcations in a signed network dynamical system}

\author{Megan Morrison$^1$, J. Nathan Kutz$^2$, and Michael Gabbay$^3$\\[.1in]
$^1$ Courant Institute of Mathematical Sciences, New York University, New York, NY 10012\\
$^2$ Department of Applied Mathematics, University of Washington, Seattle, WA 98195-3925\\
$^3$ Applied Physics Laboratory, University of Washington, Seattle WA 98105
}
\date{March 2022}

\begin{document}

\maketitle

\begin{abstract}
We investigate structural features and processes associated with the onset of systemic conflict using an approach which integrates complex systems theory with network modeling and analysis. We present a signed network model of cooperation and conflict dynamics in the context of international relations between states. The model evolves ties between nodes under the influence of a structural balance force and a dyad-specific force. Model simulations exhibit a sharp bifurcation from peace to systemic war as structural balance pressures increase, a bistable regime in which both peace and war stable equilibria exist, and a hysteretic reverse bifurcation from war to peace. We show how the analytical expression we derive for the peace-to-war bifurcation condition implies that polarized network structure increases susceptibility to systemic war. We develop a framework for identifying patterns of relationship perturbations that are most destabilizing and apply it to the network of European great powers before World War I. We also show that the model exhibits critical slowing down, in which perturbations to the peace equilibrium take longer to decay as the system draws closer to the bifurcation. We discuss how our results relate to international relations theories on the causes and catalysts of systemic war.
\end{abstract}

\keywords{social networks, political networks, complex systems, nonlinear dynamics, structural balance, community structure, polarization, critical transitions, conflict escalation, alliances}



\vspace{0.4cm}

\section{Introduction}\label{sec:intro}

Signed networks, in which positive and negative ties denote cooperative and conflictual interactions respectively, are a natural framework in which to represent militarized politics such as international relations and civil wars. Structural balance theory, which assumes that realpolitik maxims such as ``the enemy of my enemy is my friend'' are paramount in tie formation, has been the focus of applications of signed networks to international relations. This research has mostly involved tests of whether structural balance characterizes the structure of observed networks of alliances and clashes between nations \citep{maoz_what_2007, lerner_structural_2016, kirkley_balance_2019}. However, fundamental questions involving the dynamics of signed international relations have been neglected, such as how do feedbacks from competing mechanisms act to stabilize or destabilize the system leading to transitions between peace and war; what patterns of alliances and rivalries may be more susceptible to destabilization; and how does the system respond to perturbations such as the flare up of hostilities between particular countries. Models of signed network dynamics in which network tie values evolve under their mutual influence, such as the one we present here, can help illuminate such questions.

Although the most straightforward dynamical formulation of structural balance theory with continuous tie values has been successful in showing how different outcomes predicted by the theory can arise from initial conditions \citep{kulakowski_heider_2005, marvel_continuous-time_2011, morrison_community_2020}, the intent to faithfully encode structural balance theory saddles it with significant conceptual and mathematical shortcomings, limiting the ability to yield a richer and more realistic range of behaviors. The assumption that changes in tie strength between a pair of nodes are determined only by their relations with third parties ignores dyad-specific drivers of rivalry or friendship due, for instance, to territorial disputes or ideological similarities or differences. This neglect of potentially countervailing processes implies that even if the initial tie magnitudes are small they will be amplified toward a balanced end state of either systemic two-faction war or complete network harmony, whereas the international system is typically characterized by more prosaic conditions of lower level conflict and cooperation. Furthermore, once the network evolves into a state consistent with structural balance, it will stay there forever. Consequently, the absence of competing mechanisms implies that a model of pure structural balance cannot address how a system transitions from peace to war, or back again, as key conditions of the international system such as the balance of power change. Moreover, the structural balance force in the simple model is not bounded, generating infinite tie values of unlimited affinity or animosity.

We introduce a dynamical systems model of signed network evolution that includes, along with structural balance, a force that models direct dyadic interaction, thereby enabling it to capture transitions between peace and war. This direct dyadic force acts so as to pull a dyad toward a dyad-specific tie value parameter taken to represent their level of amicable or contentious relations while at peace. When structural balance pressures are absent or weak, the equilibrium tie value will be at or near this ``dyadic bias" parameter. In this way, we represent peace along a spectrum of low-level conflict and cooperation in the spirit of recent efforts to resolve peace in finer detail than simply the absence of war \citep{Diehl2016}. Along with incorporating this dyadic force, we modify the form of the structural balance force itself so that its strength saturates as mutual allies (or enemies) predominate rather than growing without bound.

As our model allows different regimes of equilibrium behavior corresponding to peace and war, it can display sharp transitions between peace and war. These transitions occur via a bifurcation, a sudden change in the qualitative nature of the solution space as a model parameter is varied \citep{strogatz_nonlinear_2016}. As the parameter controlling the sensitivity of the structural balance force to relationship changes increases relative to the dyadic force strength parameter, the system can exhibit a bifurcation from the peace state of low tie strengths to the systemic war state with strong positive and negative ties. Such an increase in the structural balance senstivity parameter reflects a systemic change that impels countries to increasingly take mutual allies or enemies into account in their relations with others; for example, due to a narrowing power gap in a rivalry between the dominant state and a rising challenger or technological developments perceived as favoring offense over defense, as have been theorized as causes of major wars \citep{VanEvera1999, copeland_origins_2000}.  The reverse transition from war to peace is also possible but occurs at a different, lower ratio of the balance to dyadic force parameters, a phenomenon known as hysteresis which reflects the underlying bistability of the system in a certain parameter regime. The capability to account for war-peace transitions in both directions distinguishes our model from many dynamics models which only address conflict onset.

We investigate the bifurcation behavior of our model in simulations and use stability analysis to derive expressions relating system parameters at the critical points where bifurcations occur. We do so for both a special case that can be reduced to a one-dimensional system and the general case involving matrix formalism. Spectral decomposition of the network into eigenvalues and eigenvectors proves to be a useful tool for illuminating the model behavior in simulations and in the stability analysis as the bifurcations between peace and war are manifested as discontinuities in the first eigenvalue of the equilibrium network. 

Community structure can be built into the system using the network of dyadic bias parameters; for instance, via a two-block structure where intra-block biases tend to be positive and inter-block ones tend to be negative. Simulations show that the peace-to-war bifurcation occurs at a lower structural balance sensitivity for networks with polarized community structure characterized by opposing factions than those without such structure. The stability analysis reveals the mathematical underpinnings of this effect: the structural balance sensitivity at which the peace-to-war bifurcation occurs decreases in inverse proportion to the first eigenvalue of the network of dyadic bias parameters, which is larger for the polarized structure case. We use this result to argue more generally that community structure in which hostile factions have a discernible signature in the eigenvalue spectrum will be more unstable to war than the case where the structure is disordered. As heightened polarization between two rival alliances has been attributed as a cause of greater instability to systemic war \citep{thompson_streetcar_2003}, our analysis therefore explains how this behavior can emerge naturally from signed network dynamics and generalizes it beyond the context of formal alliances.

Another aspect of our results bears upon the question of whether particular dyads are more destabilizing than others in terms of catalyzing major war \citep{thompson_streetcar_2003}. Standard stability analysis, as employed in obtaining the results noted above, assumes a perspective in which a parameter change causes an equilibrium to become unstable to even infinitesimal perturbations in tie values. However, it is also possible to consider how finite perturbations, such as increased cooperation or conflict among certain dyads, may destabilize the system, while keeping model parameters fixed. This can occur in the bistable state where both the peace and war states are stable and so the network will remain at its present equilibrium unless shaken out to the other one. We define an optimization framework which seeks to find the network perturbation that can trigger such a transition with the least ``energy'' in terms of the change in network tie values from their present state. We illustrate this framework using a simplified World War I context consisting of empirical networks of five great powers. We find that the ties that triggered the war require relatively low energy to destabilize the system.

We also illustrate how the network response to perturbations changes as the system nears the bifurcation point. In particular, the duration of the transient caused by the perturbation becomes longer in a power-law relationship universal to all systems that undergo a saddle-node bifurcation, as our model does, in a phenomenon known as ``critical slowing down'' \citep{kuehn_mathematical_2011}. The universality implies that critical slowing down may be useful as an early warning signal for detecting impending major state transitions in complex systems, such as the climate, ecosystems, and social systems, for which precise dynamical models are unavailable \citep{scheffer_anticipating_2012}.

Our signed network modeling approach integrates the nonlinear dynamics of complex systems \citep{strogatz_nonlinear_2016, Roberts2015} with the rich, quantitative representation of structure in complex networks \citep{newman_networks_2018, DorBatFer2020}. Accordingly, it meshes two strands of international relations research: one which casts world politics as a complex system where nonlinear feedbacks between system components can cause disproportional and dramatic responses to small changes \citep{Saperstein1999, Jervis1997, thompson_streetcar_2003, Ferguson2010, Stauffer2021} and the other which studies international relations with the concepts and methods of network analysis \citep{Maoz2011, DorGarWes2016}. The system of alliances and rivalries among states has received much attention as a cause of systemic war \citep{Snyder1984, RaslerThompson2010, LevMul2021} and our model can shed light on which patterns are particularly prone to instability, as we will illustrate by our investigation of how the peace-to-war bifurcation is affected by polarized community structure and in our application to World War I. The analytical expressions for the bifurcation conditions we derive below will greatly facilitate more generalized study of the effects of different types of community structure, both abstract and empirical, by eliminating the need to identify bifurcation points through computationally-intensive parameter sweeps of model simulations. And our finding that the interplay of the direct dyadic and structural balance forces generates a saddle-node bifurcation can guide future efforts to use universal phenomena such as critical slowing down as predictive indicators of conflict.

\section{Background}

\subsection{Dynamical systems modeling and structural balance theory in international relations}

Dynamical systems models have been useful in understanding and predicting the dynamics of complex systems in a diversity of fields such as biology, engineering, and sociology \citep{strogatz_nonlinear_2016}. While statistical models are useful in classifying and predicting phenomena, statistical models often fail to provide explicit mechanistic explanations \citep{midlarsky_preventing_1984,thompson_polarity_1986,roberts_stochastic_2016}; dynamical systems models, on the other hand, can elucidate the mechanisms responsible for the observed underlying phenomena \citep{trotta_global_2012, morrison_nonlinear_2021}. Dynamical systems models of the interaction between states or armed sub-state actors such as insurgents can help identify mechanisms for system destabilization as well as targets for reinforcing stability. These models can be used to determine the risk for minor, localized conflicts to escalate and spread extensively to other nodes in the network.  Moreover, they provide a rich set of diagnostic tools for the analysis of escalating conflicts which result in the outbreak of widespread wars.

The application of dynamical systems modeling to international relations has largely been built upon Richardson's pioneering model of arms races \citep{Gleditsch2020, Stauffer2021}. However, this stream of research predated the surge of scholarship on political networks and so did not engage with questions of network structure, although  the Richardson model itself is amenable to network formulation \citep{Ward2020}. Nor did it leverage the concepts and tools of modern bifurcation theory.

The application of network analysis to international relations has bloomed over the past two decades. Most of this work has involved unsigned networks in which either cooperative or, less frequently, conflictual ties are treated as separate networks between nations or militant groups \citep{HafKahMon2009, Maoz2011, VicMonLub2016, DorGarWes2016, ZechGabbay2016}. Signed networks, however, allow for an integration of cooperation and conflict consistent with the fact that alliances are usually formed with an eye toward a potential foe and decisions to militarily confront an opponent typically consider who might come to its aid. The investigation of world politics using signed networks has overwhelmingly been through the lens of structural balance theory \citep{cartwright_structural_1956}, a strand of research that precedes by decades the recent network turn in international relations study. In its basic form, structural balance theory is formulated on (undirected) triads and states that there are only two balanced, and hence enduring, triads: one with all positive ties and one with a single positive tie and two negative ones. In turn, these rules imply that a complete network can have only two perfectly balanced states: a harmonious state in which all ties are positive and the two hostile factions state in which the ties within a faction are positive while the ties between factions are negative. 

There has been a long-running back-and-forth about whether the international system is characterized by structural balance \citep{Harary1961, healy_balance_1973, McDonaldRosecrance1985, maoz_what_2007} with more recent work tending to support an overall tendency toward balance \citep{lerner_structural_2016, kirkley_balance_2019,burghardt_dyadic_2020}. Going beyond the question of whether the international system displays a tendency toward balance, its level of balance has been observed to fluctuate considerably over time \citep{doreian_structural_2019, burghardt_dyadic_2020}, which suggests that other forces beyond structural balance may be at work. Evidence for structural balance processes has also been found in friendship networks \citep{kirkley_balance_2019}, wild mammals \citep{ilany_structural_2013}, and in gang networks where inter-gang violence increases among gangs in imbalanced triads \citep{nakamura_violence_2020}.

\subsection{Modeling structural balance dynamics}

Given the network of initial ties, dynamical systems models can be formulated that evolve the ties over time consistent with structural balance theory. The simplest formulation evolves a symmetric network of signed edge weights, $\b{X}(t) \in \mathds{R}^{N \times N}$, as the following system of coupled, nonlinear differential equations: 
\begin{align}
    \frac{dX_{ij}}{dt} = \sum_{k=1}^N X_{ik}X_{kj}. \label{eq:kulakowski}
\end{align}
If nodes $i$ and $j$ have either both positive ties or both negative ties with a third node $k$, the product $X_{ik}X_{kj}$ will be positive and act so as to increase their tie value $X_{ij}$. In contrast, if their ties with $k$ are of opposite sign, the effect will be to decrease $X_{ij}$. Both of these effects act to increase the balance in the system. The sum determines whether the net effect of all mutual ties increases or decreases $X_{ij}$. 

Although this equation appears to have been first proposed in Lee, Muncaster and Zinnes (\citeyear{LeeMunZin1994}), it was not investigated in that paper. It was proposed independently by Kulakowski \textit{et al.} (\citeyear{kulakowski_heider_2005}) who found that in simulations this model leads to perfectly balanced outcomes consistent with the static theory. Marvel, Kleinberg, Kleinberg and Strogatz (\citeyear{marvel_continuous-time_2011}) used the matrix equation formulation of Eq.~(\ref{eq:kulakowski}),
\begin{align}
    \frac{d\b{X}}{dt} = \b{X}^2, \label{eq:marvel}
\end{align}
to obtain an analytical solution for $\b{X}(t)$. They showed that the network will generically evolve into a balanced end state as determined by the eigenvector of the initial adjacency matrix $\b{X}(0)$ having the most positive eigenvalue, which grows the fastest. If this first eigenvector consists of all positive components, then the system will evolve into the harmonious state. On the other hand, if the first eigenvector has components of both positive and negative sign, then the system will evolve into the two hostile factions state. Marvel \textit{et al.} (\citeyear{marvel_continuous-time_2011}) used this model to show good agreement with the alliances which formed in World War II and the split of the Zachary karate club. Morrison and Gabbay (\citeyear{morrison_community_2020}) extended this model to initial conditions containing community structure generated by a two-block stochastic model. They found that phase transitions in the initial network structure determine whether the two final factions align with the block structure or instead have random memberships unrelated to the blocks or, alternatively, whether it is the harmonious state that emerges.

Equation~(\ref{eq:kulakowski}) can be analyzed for the case of a directed network in which case the system evolves to four factions instead of two from random initial conditions \citep{veerman_social_2018}. It is also possible to define a variant model for directed networks  that replaces $X_{ik}X_{kj}$ on the right-hand side of (\ref{eq:kulakowski}) with $X_{ik}X_{jk}$, which can be shown to evolve into two factions for random initial conditions \citep{traag_dynamical_2013}. Although our focus here is on deterministic dynamical systems, structural balance theory can also be implemented as a stochastic model \citep{antal_social_2006}.

As we employ a control theory framework to investigate the patterns of network perturbations that are most destabilizing, we note some previous work applying control theory to structural balance. Gao \textit{et al.} (\citeyear{gao_structural_2018}) has investigated the origins of balancing forces with respect to nodal interactions, and has considered how to control structural balance edge dynamics using node dynamics, with the assumption that the dynamics of nodes and edges are interdependent. Wongkaew \textit{et al.} (\citeyear{wongkaew_control_2015}) has considered how to control structural balance dynamics by introducing a `leader' to the system, while Summers and Shames (\citeyear{summers_active_2013}) have evaluated the control abilities of existing nodes in a network by changing ties.

While the simple model of Eq.~(\ref{eq:kulakowski}) has yielded great insight into structural balance dynamics and will serve as the main point of departure, it also has important shortcomings. Although it is able to determine the outcomes that emerge from a network solely under structural balance dynamics, it has no equilibrium, evolving ties toward positive or negative infinity (and does so in finite time \citep{marvel_continuous-time_2011}). Once the model is turned on, the network perforce evolves into a state of complete war or harmony and so it cannot capture the stable, pre-escalatory state which characterizes the international system for long stretches of time before and after systemic wars. Its empirical application therefore depends upon the assumption that structural balance dynamics are either off or on and so, while useful for predicting the composition of the opposing sides once the war starts, the model is silent about what patterns of relationships might set off such a war in the first place. Nor can the model describe de-escalation dynamics, from systemic war back to a peaceful state since it inherently seeks to increase balance. Accordingly, we aim to build upon the dynamical systems model of conflict escalation in Eq.~(\ref{eq:kulakowski}) to create a model that can manifest a wider range of behaviors in the international system --- stability in a peaceful, low-conflict state, transitions to systemic war, and de-escalation.

\section{Signed network model of cooperation and conflict dynamics} \label{sec:signed_model}

We present a model for signed network edge dynamics that incorporates two major forces: (1) a dyadic force that seeks to stabilize a dyad's tie value at a low level of cooperation or conflict as determined by the particular history and context of their relationship; and (2) a bounded structural balance force, rather than the unbounded force of Eq.~(\ref{eq:kulakowski}). In addition, we will allow for transient perturbations of particular tie values to allow for investigation of how conflict flareups or temporary cooperation may affect the stability of the whole system. Our model contains a low-level equilibrium that corresponds to conventional peace in the international system, a high-conflict equilibrium that corresponds to systemic war, and a high-cooperation equilibrium corresponding to global harmony. Bifurcations and perturbations to the system can induce the network to transition between these states. Understanding how these bifurcations, particularly those between peace and war, depend on the relative strengths of the parameters associated with the forces in the model will be the primary focus of our analysis of the model.

In our model, the dynamics governing the rate of change for edge $X_{ij}(t)$ between nodes $i$ and $j$ in a network containing $N$ nodes is given by 
\begin{align} \label{eq:dyn_sys_edge_ij_full}
    \frac{dX_{ij}}{dt} = -\beta (X_{ij} - X_{Dij}) + L \tanh \left( \frac{1}{\gamma} \sum_{k=1}^N X_{ik}X_{kj} \right) + X_{Tij}(t),
\end{align}
where $-\beta (X_{ij} - X_{Dij})$ is the direct dyadic force, $ L \tanh (\gamma^{-1} \sum_{k=1}^N X_{ik}X_{kj})$ is the bounded structural balance term, and $X_{Tij}(t)$ represents fast-timescale perturbations. Although the above equation can be applied to directed networks, we will only consider the undirected case here so that $X_{ij}=X_{ji}$. The other matrix quantities, $X_{Dij}$ and $X_{Tij}$, are similarly symmetric.

The strength of the direct dyadic force is scaled by the parameter $\beta$, which we take to be the same for all dyads.\footnote{Technically, $\beta$ can be removed from (\ref{eq:dyn_sys_edge_ij_full}) by rescaling $t \rightarrow \beta t$ and $L \rightarrow L/\beta$ (ignoring the perturbation term), which implies that $L$ and $\beta$ should only enter the dynamics via their ratio. However, $\beta$ is needed if one wants $t$ to correspond to a conventional unit of time.} We refer to $X_{Dij}$ as the dyadic bias parameter, which can take on dyad-specific values. Under the action of the direct dyadic force alone, $X_{ij} \rightarrow X_{Dij}$ so that $X_{Dij}$ is the equilibrium tie value. A weak structural balance force will cause only a small shift in the equilibrium values from the $X_{Dij}$ but, as will be seen, a sufficiently strong structural balance force will induce a transition to a high level of equilibrium tie values. Accordingly, the dyadic bias parameters correspond to relatively low levels of cooperation and conflict. There are many factors that can affect the propensity for two states to cooperate such as mutual trade interests or common ideology. Other factors such as common territorial aspirations or disparate ideologies predispose discordant states to hostility. We will typically, but not always, use the bias parameters as the initial tie values in our simulations, $X_{ij}(0)=X_{Dij}$.

The bounded form of the structural balance force is produced by the S-shaped hyperbolic tangent function, which asymptotes to $\pm L$ as its argument tends to $\pm\infty$. $L$ therefore scales the maximum magnitude of the structural balance force. The parameter $\gamma$ is the characteristic half-width of the transition region between the negative and positive plateaus. It will be convenient for subsequent analysis to define the structural balance sensitivity, $\alpha$, which is the characteristic slope of the transition region, $\alpha=L/\gamma$. In a small neighborhood around zero, the hyperbolic tangent is approximately linear so that the structural balance force is approximately given by $\alpha \sum_{k=1}^N X_{ik}X_{kj}$ and so reduces to the unbounded form of structural balance in Eq.~(\ref{eq:kulakowski}) in this neighborhood (apart from $\alpha$, which is unnecessary in (\ref{eq:kulakowski})).  The strength of the balance force is also affected by the size of the network $N$, and the tie density. Either increasing the size or the density of the network tends to increase $\sum_{k=1}^N X_{ik}X_{kj}$.

The term $X_{Tij}(t)$ contains transient perturbations to the system caused by fast-timescale events such as a militarized conflict between two states or cooperation against an adversary. It could take the form of an impulse, positive or negative, to a given dyad or a set of dyads. These impulses may knock the system out of its current equilibrium. Note that the $X_{Tij}(t)$ correspond to finite perturbations as distinguished from the infinitesimal perturbations always assumed to be present and which prevent the system from staying in unstable equilibria.

We note that the dynamics of this model results in nonzero self-ties,  $X_{ii}$. These ties will always be positive because the structural balance term becomes a sum over $X^2_{ik}$ (unless one makes the odd choice of a negative self-bias, $X_{Dii}<0$, implying that the node has a conflictual relationship with itself). Since it is not obvious how to empirically measure a self-tie, they are often set to zero. Yet, conceptually, a positive self-tie is not problematic as it can be interpreted as self-cooperation. When a node is a composite entity such as a country, a high self-cooperation, for instance, could correspond to greater mobilization of the populace during a war. Whether or not the self-tie is set to zero would be simply a conceptual choice were it not for the fact that it does enter into the dynamics of the other ties, $X_{ij}$. In the roughly linear region of the hyperbolic tangent, the self-ties contribute a force $\alpha (X_{ii} + X_{jj})X_{ij}$, which has the effect of adding a nonlinear term to the direct dyadic force. Given the positivity of the self-ties, this term produces a tendency for a dyad to be further pushed in the direction of its present sign, toward more cooperation for $X_{ij}(t)>0$ and more conflict if $X_{ij}(t)<0$, corresponding to mutually reinforcing behavior between the nodes in the dyad.\footnote{Such mutually reinforcing behavior was included in the model of Lee, Muncaster, and Zinnes (\citeyear{LeeMunZin1994}). They added a dyadic force proportional to $+X_{ij}$ to Eq.~(\ref{eq:kulakowski}) so that friendship breeds greater friendship and likewise for hostility. But this dyadic force also causes tie values to blow up as does (\ref{eq:kulakowski}) rather than the stabilization to equilibrium produced by the linear restoring dyadic force appearing in (\ref{eq:dyn_sys_edge_ij_full}).} However, this term is just one of $N$ terms appearing in the structural balance force and so becomes ever more negligible as the number of nodes grows larger. 

\section{Model behavior}

In this section, we use simulations to explore the dynamics of the above model. The model shows a peace-to-war bifurcation as the structural balance sensitivity $\alpha$ is increased. which reflects the feed-forward escalatory conflict that spreads throughout the entire network, requiring each node to participate in the global conflict and take one of two sides. Once the network has stabilized at a high-conflict state, it requires a different bifurcation, which occurs at a lower $\alpha$ value, to bring the system back to a peaceful state. We illustrate how community structure affects the bifurcation threshold and how the bifurcation is manifested as a discontinuity in the equilibrium network eigenvalue spectrum and in a measure of its balance level. The existence of a bifurcation directly between the war and harmonious states is also observed which, however, does  not display a discontinuity in the eigenvalue spectrum.

\subsection{Bifurcations in simulation}

 \begin{figure}[t]
    \centering
    \includegraphics[width=0.85\linewidth]{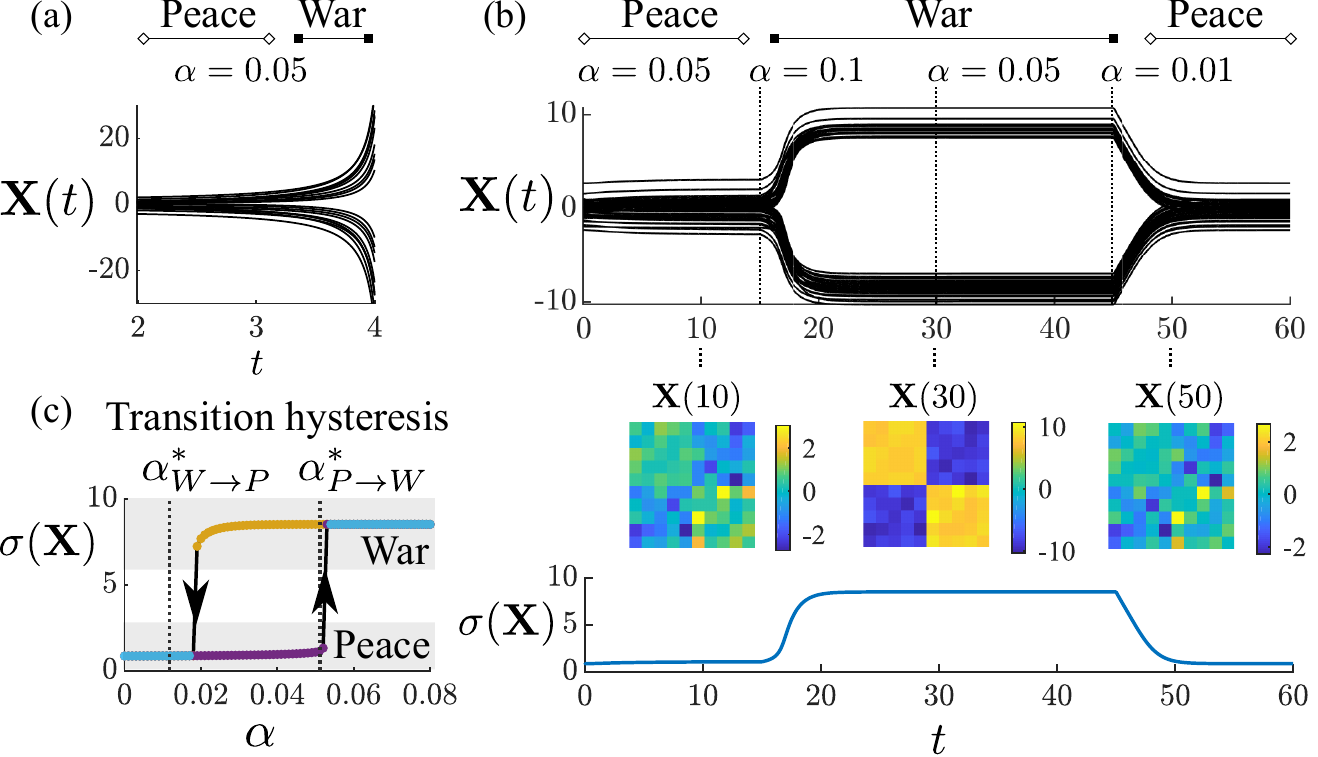}
    \caption{Simulations of signed network model dynamics. (a) Time series of tie values $X_{ij}(t)$ for pure structural balance model, Eq.~(\ref{eq:marvel}), showing unbound dynamics. (b) Time series for the dyadic force and structural balance model, Eq.~(\ref{eq:dyn_sys_edge_ij_full}), as the balance sensitivity $\alpha$ is changed at the times indicated by the dotted lines; $\beta = 1$, $L=8$, $X_{Tij}=0$, and $X_{ij}(0)=X_{Dij}$. $X_{Dij} \sim 0.8 \mathcal{N}(0,1) +0.4 \b{v}\b{v}^T$ is a random symmetric matrix with community structure, $\b{v} = [\b{1}; -\b{1}] \in \mathds{R}^N$. Snapshots of network ties are taken at $t=10$, $30$, and $50$. The standard deviation of network ties, $\sigma(\b{X})$, is measured over time. (c) Standard deviation of network ties as a function of the structural balance parameter $\alpha$ as well as the initial state (war or peace). The vertical dotted lines are the predicted critical values of $\alpha$ for the peace-to-war and war-to-peace bifurcations from Eqs.~(\ref{eq:alpha_PW_general}) and (\ref{eq:alpha_WP_gen}) respectively.
    }
    \label{fig:intro_fig}
\end{figure}

Figure~\ref{fig:intro_fig}(a) and (b) compare the tie dynamics for a network governed by pure structural balance dynamics (Eq.~\ref{eq:kulakowski}) versus a network governed by the simultaneous operation of the direct dyadic and structural balance forces (Eq.~(\ref{eq:dyn_sys_edge_ij_full}) with $X_{Tij}=0$). As noted above, the pure structural balance dynamics results in ties that blow up to positive or negative infinity. This phenomenon occurs for all initial conditions with no means of reverting back to a peaceful state with low tie values.

Figure~\ref{fig:intro_fig}(b) shows the dynamics of Eq.~\ref{eq:dyn_sys_edge_ij_full} in which the structural balance sensitivity, $\alpha = L/\gamma$, is modified at times $t=15$, $30$, and $45$ (by changing $\gamma$). The dyadic bias parameters, which also serve as the $t=0$ tie values, are randomly set but have an underlying two-block community structure. At the parameter shift times, the network state is not reset to $\b{X}_D$, but rather the present tie values serve as the initial conditions for the new $\alpha$ value and the system is allowed time to reach equilibrium. 

The first time segment from $t=0$ to $t=15$ with $\alpha=0.05$ shows that the system quickly reaches an equilibrium in which the tie values are close to their initial $X_{Dij}$ ones. Thus, unlike pure structural balance dynamics, our model possesses a stable state spanning a range of relatively small, positive and negative tie values as seen in the low standard deviation of the ties in the bottom plot of Fig.~\ref{fig:intro_fig}(b). We refer to this state as the peace equilibrium. It accords with recent work that seeks to provide a finer-grained characterization of peace as shades of rivalry and friendship, rather than as an undifferentiated state of simply ``not war'' as has often been assumed in conflict studies \citep{Diehl2016}.

In the second time segment from $t=15$ to $t=30$, $\alpha$ is doubled to $0.1$ and the peace state destabilizes and the system transitions to the high-conflict, systemic  war state. The ties diverge into two bunches of strong positive and negative tie values, yielding a much larger standard deviation than the peace state. The increase in structural balance also polarizes the network into the two blocks underlying the dyadic bias parameters. At $t=30$, $\alpha$ is lowered back down to $0.05$, its value in the first time segment. Yet, this does not bring the system back to the peace state. $\alpha$ must be decreased even further, to $0.01$ at $t=45$, to induce the transition back to the low-deviation peace state. This behavior is an example of hysteresis, where a system's future state is determined not only by its parameter values but also by its current state \citep{strogatz_nonlinear_2016}.

Figure~\ref{fig:intro_fig}(c) shows the tie standard deviation, $\sigma(\b{X})$, for various $\alpha$ values and initial conditions. For $\alpha<0.018$ only the peace state is stable and the network will converge to a peaceful state for all initial conditions. For $\alpha \in (0.018, 0.053)$, both the peace and war states are stable and the system is bistable as is consistent with hysteresis; which of the two equilibria the network converges to is dependent on initial conditions. For $\alpha>0.053$ only the war state is stable and the network will converge to it for all initial conditions. 

\subsection{Bifurcations in the eigenvalue spectrum}

It will be helpful to analyze the network's adjacency matrix and its dynamics in terms of its eigenvalues and eigenvectors. The set of eigenvectors $\b{s}_i$ of the real, symmetric matrix $\b{X}$ satisfy the following equation:
\begin{align}
	\b{X} \b{s}_i = \lambda_i \b{s}_i,
\end{align}
where the eigenvalues $\lambda_i$ are real and ranked in order of descending value. The eigenvectors $\b{s}_i$ form an orthonormal set. In matrix form, the eigenvector decomposition can be written as
\begin{align}
    \b{X} = \b{S} \Lambda \b{S}^T,
\end{align}
where the columns of $\b{S}$ are the $\b{s}_i$ and $\Lambda$ is a diagonal matrix containing the corresponding eigenvalues $\lambda_i$. The eigenvectors form an alternative, and often more intuitive, coordinate basis for $\b{X}$. In unsigned networks, for example, the first eigenvector components are proportional to the node eigenvector centralities whereas, for signed networks, strong two-faction structure will be reflected in the first eigenvector in which the factional memberships are of opposite sign.

\begin{figure}[t]
    \centering
    \includegraphics[width=0.85\linewidth]{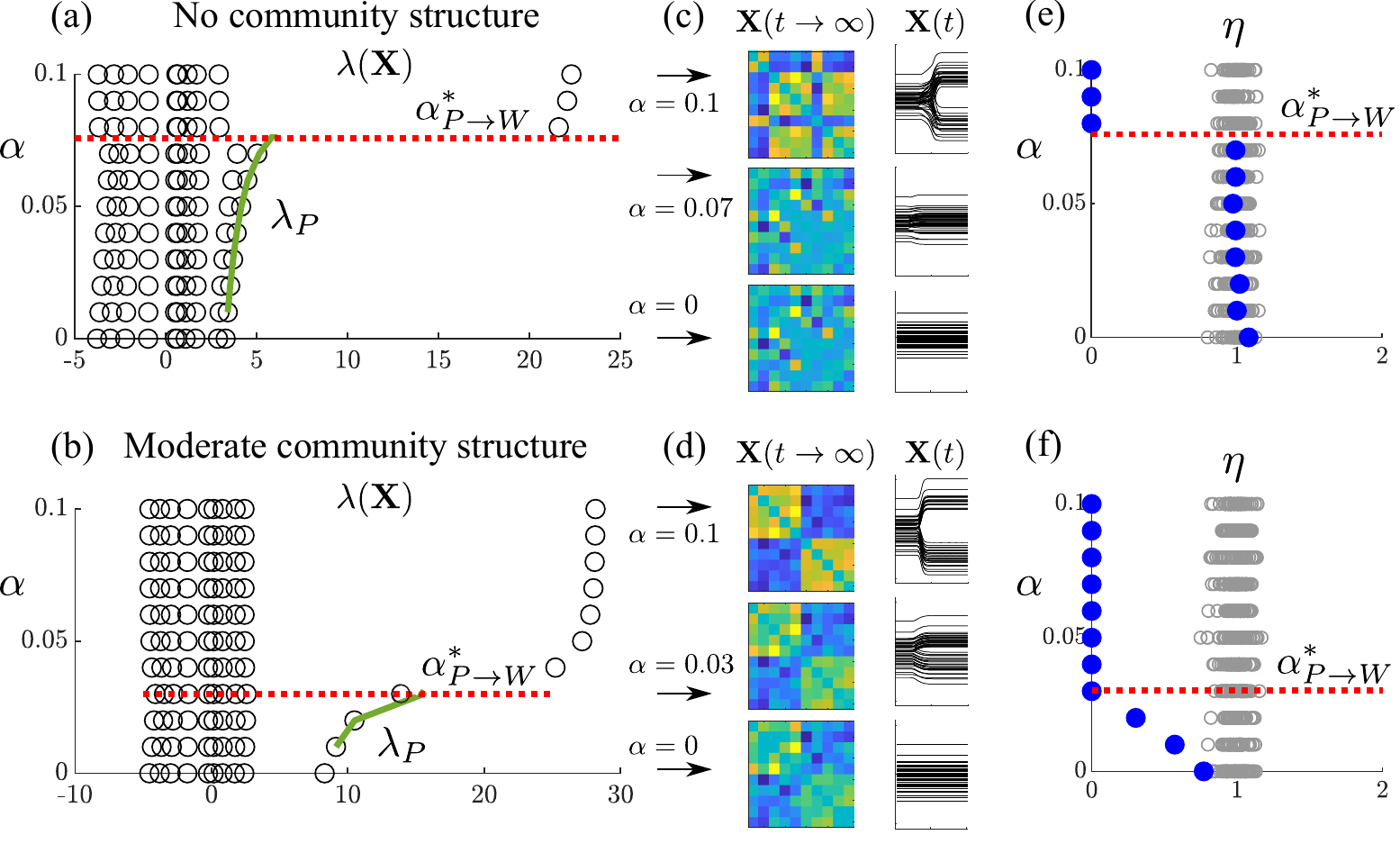}
    \caption{Eigenvalues, $\lambda_i$, and balance levels, $\eta$, of the network at equilibrium for increased balance sensitivity $\alpha$. $L=2$, $\beta=1$. (a) Stable state eigenvalues as a function of $\alpha$ for a random $\b{X}_D \sim 0.8 \mathcal{N}(0,1)$ containing no community structure. Theoretically computed bifurcation value $\alpha_{P \rightarrow W}^*$ (red dashed line). (b) Stable state eigenvalues for a network with $\b{X}_D\sim 0.8 \mathcal{N}(0,1) +0.4 \b{v}\b{v}^T$
    containing community structure, $\b{v} = [\b{1}; -\b{1}] \in \mathds{R}^N$. (c) Networks at equilibrium and edge dynamics for $\alpha = 0.1$, $\alpha = 0.07$, and $\alpha = 0$ for $\b{X}_D$ without community structure. (diagonals set to zero) (d) Networks at equilibrium and edge dynamics for $\alpha=0.1$, $\alpha=0.03$, and $\alpha=0$ for $\b{X}_D$ with community structure. (e) Balance levels $\eta$ of the equilibrium network (blue circles) as a function of $\alpha$. Null model simulations shown in gray. (f) Balance levels for a network with a random $\b{X}_D$ containing community structure. The green curves in (a) and (b) show the analytical expression, (\ref{eq:lambda_P}), for the first eigenvalue in the peace state.}
    \label{fig:eigvals_alpha}
\end{figure}

Figure~\ref{fig:eigvals_alpha} compares the simulation results in which the network of dyadic biases $\b{X}_D$, is either completely random with no structure at all or contains community structure.  Community structure can be generated and studied using stochastic block models (Sec.~\ref{sec:commstruc}). The dyadic biases also serve as the initial conditions $\b{X}(0)=\b{X}_D$ for each simulation run. In  Fig.~\ref{fig:eigvals_alpha}(a), we plot the adjacency matrix spectrum as $\alpha$ is increased. Each horizontal slice contains all the eigenvalues of $\b{X}$ at equilibrium for a given $\alpha$ value.  The peace-to-war bifurcation occurs at the critical value of the $\alpha$ parameter, $\alpha^*_{P \rightarrow W}$, after  which conflict escalates through the system. For  $\alpha<\alpha^*_{P \rightarrow W}$, the system is stable in the peace state and the eigenvalues form a dense band although the first two eigenvalues show some growth as $\alpha$ is increased. For $\alpha>\alpha^*_{P \rightarrow W}$, the peace state is destabilized and the system transitions to the war state.  The leading eigenvalue is now much larger than the others, making the equilibrium adjacency matrix close to rank 1. The curves in Figure~\ref{fig:eigvals_alpha}(a) show that the analytically computed leading eigenvalue in the peace state (Eqs.~(\ref{eq:lambda_P}), derived in the Appendix) is in good agreement with the simulation results.  Figure~\ref{fig:eigvals_alpha}(c) visualizes the equilibrium adjacency matrix at three $\alpha$ values as well as the corresponding time series of the tie weights. At low $\alpha$ values, the network ties are close to the random states in $\b{X}_D$, while beyond $\alpha^*_{P \rightarrow W}$ the final network is divided into two camps with positive ties within each group and large negative ties between the two groups (the nodes could be reordered so they appear as blocks).

The spectrum of the case where $\b{X}_D$ contains community structure is shown in Fig.~\ref{fig:eigvals_alpha}(b). For $\alpha=0$, there is no structural balance force and the tie values remain at their dyadic biases as seen in the time series of Fig.~\ref{fig:eigvals_alpha}(d), so the equilibrium $\b{X}$ spectrum is the same as $\b{X}_D$. Unlike the no structure case, the first eigenvalue is substantially larger than the rest reflecting the underlying two-block structure of $\b{X}_D$, which can be seen in the  $\alpha=0$ matrix visualization. This community structure results in a much smaller $\alpha^*_{P \rightarrow W}$ than the no structure case. This suggests that networks with clear community structure, strong enough to appear in the first eigenvalue may be more easily destabilized than those without. We present a more general argument about this behavior in Sec.~\ref{sec:commstruc}.

Balance levels in signed networks can be measured by global balance metric $\eta$ from Kirkley \textit{et al.} (\citeyear{kirkley_balance_2019}) where $\eta = 0$ corresponds to perfect balance and $\eta = 1$ corresponds to the average level of balance generated in null model simulations. The no structure case does not show a significant level of balance until after the bifurcation (Fig.~\ref{fig:eigvals_alpha}(e)) at which point the network transitions to the war state and becomes perfectly balanced. The network with initial community structure, in contrast, shows significant levels of balance as compared with the null model values that increase with $\alpha$ even before the bifurcation (Fig.~\ref{fig:eigvals_alpha}(f)). These plots show that the onset of war greatly increases the level of structural balance in the network. Yet, the community structure plot also reveals that it is possible for the network to be balanced in a statistical sense in the peace state. That balance can characterize peace and increases due to major war has been observed empirically for the international system \citep{kirkley_balance_2019}.

The peace-to-war bifurcation threshold is given by
\begin{align}
    \alpha^*_{P\rightarrow W} = \frac{\beta}{4 \lambda_{D1}}, 
    \label{eq:alpha_PW_general}
\end{align}
where $\lambda_{D1}$ is the leading eigenvalue of $\b{X}_D$. We derive this bifurcation in Appendix~\ref{app:general_case}. The values for $\alpha^*_{P\rightarrow W}$ predicted by this formula are in good agreement with the observed thresholds as shown by the red dashed lines in Fig.~\ref{fig:eigvals_alpha}(a) and (b). The community structure case has a higher $\lambda_{D1}$ than the no structure case as seen in the $\alpha=0$ spectra, which by (\ref{eq:alpha_PW_general}) lowers the bifurcation threshold.

Even without the derivation, Eq.~(\ref{eq:alpha_PW_general}) can be motivated intuitively. That increasing the the direct dyadic force strength $\beta$ increases $\alpha^*_{P\rightarrow W}$ makes sense as the dyadic force seeks to maintain the peace state of low conflict and cooperation. When the elements of $\b{X}_D$ are drawn from a zero mean distribution as in Fig.~\ref{fig:eigvals_alpha}, the  squared eigenvalue, $\lambda_{D_i}^2$, is proportional to the variance of $\b{X}_D$ carried by each eigenvector.\footnote{This is because the eigenvectors of $\b{X}_D^2$ (proportional to the covariance matrix) are the same as $\b{X}_D$ but the eigenvalues are squared.} The appearance of $\lambda_{D1}$ in the denominator of (\ref{eq:alpha_PW_general}) then reflects the fact that networks with greater values of $\lambda_{D1}$ are more polarized in the peace state and so more unstable to war.

\begin{figure}[t]
    \centering
    \includegraphics[width=0.85\linewidth]{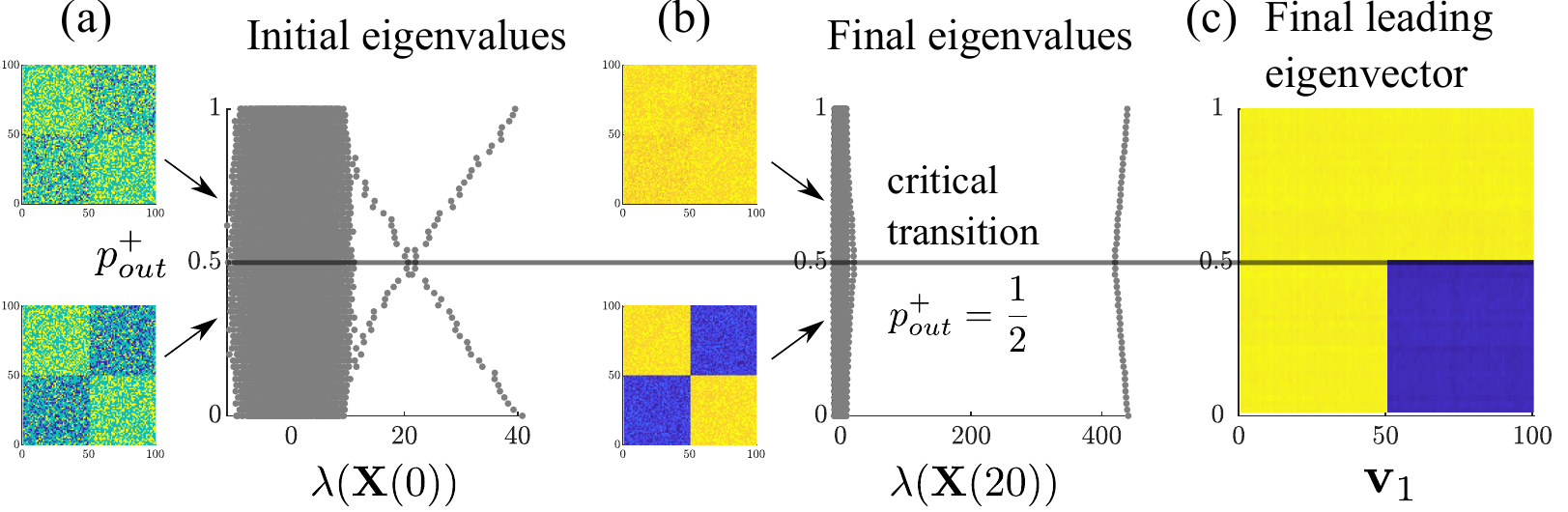}
    \caption{Simulation illustrating war-to-harmony bifurcation. (a) Eigenvalues of $\b{X}_D$, also equivalent to initial network $\b{X}(0)$, for varying parameter value $p_{out}^+$. Sample matrices at left depict how  the ties are more positive and community structure is weaker above the crossing at $p_{out}^+=0.5$ than below. (b) Eigenvalues of the equilibrium network after simulation of Eq.~(\ref{eq:dyn_sys_edge_ij_full}) with $N=100$, $\beta = 1$, and $\alpha = 0.02$. (c) Components of leading eigenvector of equilibrium network. Other stochastic block model parameters: $d_{in}=d_{out} = 0.4$, $p_{in}^+=1$.}
    \label{fig:homog_fact}
\end{figure}

We note that for sufficient net positivity in $\b{X}_D$, a bifurcation between the peace and global harmony states occurs as the balance sensitivity $\alpha$ is increased, analogous to the peace-to-war transition. It is also possible to have a direct transition between the war and harmony states. In the pure structural balance dynamics model, Eq.~(\ref{eq:kulakowski}), the war-to-harmony transition occurs when strong two-faction community structure becomes subordinate to harmonious one-faction structure in the initial network \citep{morrison_community_2020}. We consequently expect our model (\ref{eq:dyn_sys_edge_ij_full}) to likewise undergo this bifurcation for sufficiently strong structural balance and initialized network structure. To illustrate this bifurcation, instead of varying $\alpha$, we will vary the dyadic bias parameters via the outgroup affinity $p_{\text{out}}^+$, the (conditional)  probability of forming a positive tie with the outgroup in the stochastic block model which can be used to generate $\b{X}_D$ (see Sec.~\ref{sec:commstruc}). In Fig.~\ref{fig:homog_fact}(a), the reversal of dominant structure is observed as the crossing in the spectrum of $\b{X}_D$, which plays a double role as the initial network. The spectrum of the final network, however, does not display a crossing; nor is there a discontinuity in the first eigenvalue as in Fig.~\ref{fig:eigvals_alpha}. Instead, it is the components of the first eigenvector which manifest the discontinuous signature of the bifurcation (Fig.~\ref{fig:homog_fact}(c)). The block structure of the eigenvector components below $p_{\text{out}}^+=0.5$ yields the two-faction final network (Fig.~\ref{fig:homog_fact}(b)), whereas the uniformly positive components above $p_{\text{out}}^+=0.5$ yields the global cooperation of the harmony state.

\section{Stability analysis}

We now provide a mathematical explanation underlying the bifurcations seen in the simulations. First, we treat a simple special case that can be reduced to a one-dimensional system, yet illustrates the central aspects of the bifurcation behavior. We determine approximate formulas for the equilibria, both stable and unstable, in our dynamical system and use these to predict the critical value of the structural balance sensitivity at which the bifurcation occurs.  We then summarize the results for the general case, which are derived in the Appendix.

\subsection{Special case: equal and opposite factions} \label{sec:specialcase}

Consider the special case where there are two equally sized, opposing factions in which all dyadic bias parameters have the same magnitude $\mu$, but are positive for ties within the same faction and negative between factions so that $X_{Dij}=\pm \mu$. We set the initial condition to be $X_{ij}(0)=X_{Dij}$. Equation~(\ref{eq:dyn_sys_edge_ij_full}) then reduces to two equations,
\begin{align}
    \frac{dx}{dt} &= -\beta(x-\mu)+L \tanh(\frac{\alpha N}{L}x^2) \label{eq:ingroup_ties}\\
    \frac{dy}{dt} &= -\beta(y+\mu)-L \tanh(\frac{\alpha N}{L}y^2), \label{eq:outgroup_ties}
\end{align}
for ingroup ties $x$ and outgroup ties $y$, with initial conditions, $x(0)=\mu$ and $y(0)=-\mu$. However, if we substitute $y=-x$ in the bottom equation, we recover the top one and its initial condition, showing that the two equations are not independent. The system therefore reduces to one equation as solving for $x$ using (\ref{eq:ingroup_ties}) then determines $y$; the conflictual inter-faction ties are simply the negative of the cooperative intra-faction ties.  

\begin{figure}[t]
    \centering
    \includegraphics[width=0.85\linewidth]{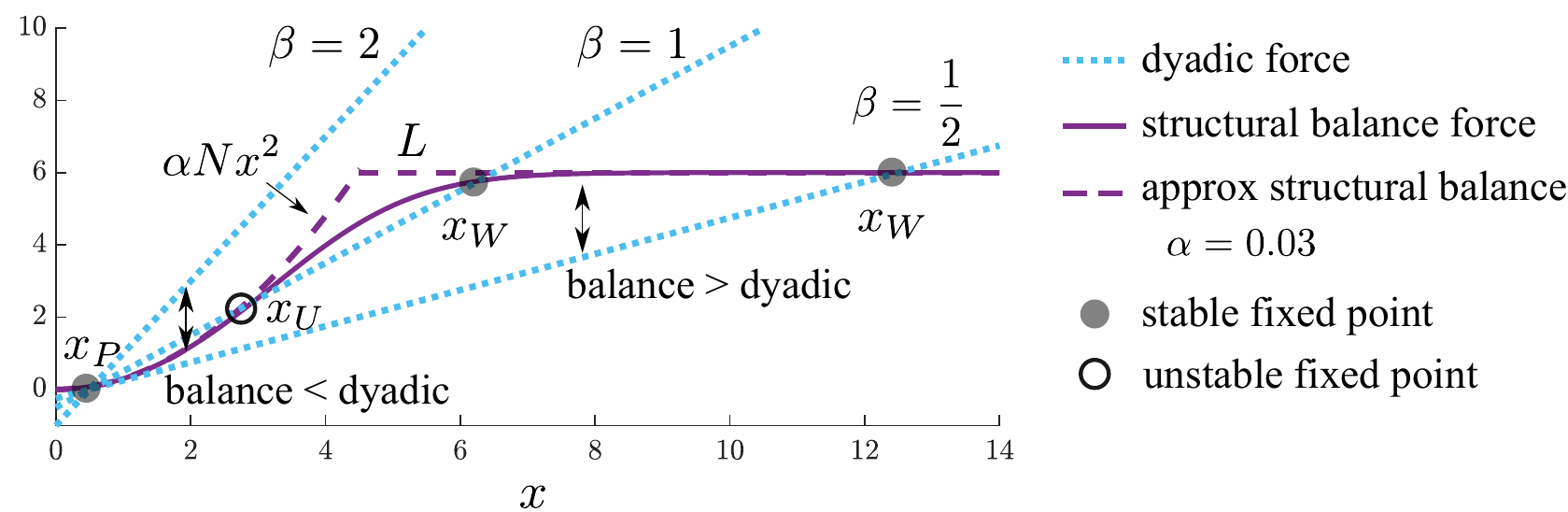}
    \caption{Strength of structural balance (purple) and direct dyadic (blue) forces and behavior of the equilibria in the special case. The approximations of the structural balance force (dashed) used in (\ref{eq:special_pwlin_top}) and (\ref{eq:special_pwlin_bot}) agree very well with the exact form  (\ref{eq:ingroup_ties}) (solid), but are less accurate in the shoulder region.  Stable (solid circles) and unstable (open circles) fixed points occur at intersections of balance and dyadic forces. $\alpha=0.03$, $L=6$, $N=10$, and $\mu = 1/2$.}
    \label{fig:special_case_mikes_way}
\end{figure}

We make Eq.~(\ref{eq:ingroup_ties}) more tractable by replacing  the hyperbolic tangent term on the positive axis with a piecewise linear approximation consisting of a line of unit slope connecting the origin to a plateau of height $L$ such that: $L \tanh \theta \rightarrow \theta$ for $0 \leq\theta < L$ and $L\tanh\theta \rightarrow L$ for $\theta \geq L$, where $\theta=\alpha Nx^2$. This approximation for the structural balance term is shown by the purple dashed line in Fig.~\ref{fig:special_case_mikes_way} (note it is piecewise linear in $\theta$ not $x$, which is why the transition section is concave upwards). Under this approximation, Eq.~(\ref{eq:ingroup_ties}) becomes
\begin{align}
	\frac{dx}{dt} &= -\beta(x-\mu)+ \alpha Nx^2, & 0 \leq x < \sqrt{\frac{L}{\alpha N}}  \label{eq:special_pwlin_top}	\\ 
	\frac{dx}{dt} &= -\beta(x-\mu)+ L, & x \geq \sqrt{\frac{L}{\alpha N}}. \label{eq:special_pwlin_bot}
\end{align}
Using these equations, we can derive simple formulas for the fixed points in the system: the peace and unstable fixed points will be obtained from (\ref{eq:special_pwlin_top}) and the war fixed point from (\ref{eq:special_pwlin_bot}).

The equilibria of the above equations are found by setting $dx/dt=0$, which implies that the direct dyadic and structural balance forces are of equal magnitude but of opposite sign. The right-hand side of Eq.~(\ref{eq:special_pwlin_top}), being quadratic, has two roots. One yields the peace equilibrium, which is stable and occurs at
\begin{align}
    x_P = \frac{\beta - \sqrt{\beta(\beta - 4 N \alpha \mu)}}{2 N \alpha}.
    \label{eq:xp_special}
\end{align}
The other fixed point is unstable and occurs at
\begin{align}
    x_U = \frac{\beta + \sqrt{\beta(\beta - 4 N \alpha \mu)}}{2 N \alpha}.
    \label{eq:xu_special}
\end{align}
Setting the right-hand side of Eq.~(\ref{eq:special_pwlin_bot}) to zero, yields the fixed point corresponding to the war state, which is stable:
\begin{align}
    x_W = \frac{L}{\beta} + \mu.
    \label{eq:xw_special}
\end{align}
The above fixed points refer to the ingroup tie values --- the outgroup tie values are their negatives. We observe that structural balance enters into these fixed points differently, for the war state via its maximum value, $L$, whereas for the peace and unstable states via the sensitivity $\alpha$, which affects how much the structural balance force grows in response to changes around its neutral point.

Figure~\ref{fig:special_case_mikes_way} shows how the nature of the equilibrium solutions changes as the dyadic force strength parameter $\beta$ is varied while the structural balance parameters remain constant. The fixed points are located at the intersections of the linear dyadic force and the nonlinear structural balance force, where their magnitudes are equal but point in opposite directions with the dyadic force acting to decrease $x$ (toward  $\mu$) and the balance force acting to increase $x$. The stability of a fixed point with respect to small perturbations is determined by the relative magnitudes of the forces on either side of it.  At moderate levels of the dyadic strength, $\beta = 1$, there are three fixed points in the system. The high cooperation (and conflict) war equilibrium at $x_W\approx 6$ is stable since perturbations to its left (right) will be pulled back by the stronger balance (dyadic) force. The peace equilibrium at $x_P \approx 0.6$ is similarly stable. But the intermediate fixed point, $x_U$ is unstable since the stronger balance force pulls positive perturbations away from $x_U$ and towards $x_W$ and negative perturbations are drawn toward $x_P$ by the stronger dyadic force so that $x \rightarrow x_W$, when $x> x_U$ and $x \rightarrow x_P$, when $x< x_U$. When the dyadic strength is strong, $\beta = 2$, the stable peace state in the only fixed point, so that $x \rightarrow x_P$ for all $x$. Finally, when the dyadic stability is weak, $\beta = 1/2$, the stable war state at $x_W \approx 12.5$ is the only fixed point so that $x \rightarrow x_W$ for all $x$ (although it is difficult to discern in the figure, the dyadic and balance forces do not intersect at low $x$).

The bifurcation responsible for the changes between the different solution regimes is a saddle-node bifurcation. Two fixed points disappear when a stable fixed point collides with the unstable one as a parameter is changed. For example, if we increase $\beta$ starting at from the $\beta=1$ condition of  Fig.~\ref{fig:special_case_mikes_way}, the intersections determining $x_U$ and $x_W$ will draw closer together until they coincide when the dyadic force is tangent to the shoulder of the exact structural balance force (or intersects the corner of the approximate one). This coincidence point marks the war-to-peace bifurcation as it disappears upon further increase in $\beta$, leaving only the $x_P$ fixed point. Analogously, the war-to-peace bifurcation occurs when $x_P$ and $x_U$ collide, leaving only $x_W$.

We can now derive expressions relating the parameters at the bifurcation points. The peace-to-war bifurcation occurs when $x_P = x_U$. Equating (\ref{eq:xp_special}) and (\ref{eq:xu_special}) implies that the square root term must vanish, which yields the condition (ignoring the $\beta=0$ solution, which is simply pure structural balance),
\begin{align}
\alpha^*_{P\rightarrow W} = \frac{\beta}{4 N \mu}, 
\label{eq:alpha_PW_special}
\end{align}
where we have solved for the critical value of the structural balance sensitivity. Similarly, the war-to-peace bifurcation occurs when $x_U=x_W$. Equating (\ref{eq:xu_special}) and (\ref{eq:xw_special}) yields
\begin{align}
\alpha^*_{W\rightarrow P} = \frac{L}{N(\mu+L/\beta)^2}.
\label{eq:alpha_WP_special}
\end{align}
Equation~(\ref{eq:alpha_PW_special}) shows that the peace-to-war transition occurs at larger $\alpha$ values for larger dyadic force strengths, $\beta$, a behavior that accords with the dyadic force's role in stabilizing the peace equilibrium.  Increasing the dyadic bias parameter, $\mu$, destabilizes the system at a lower $\alpha^*_{P\rightarrow W}$, signifying that greater polarization in the peace state makes it more susceptible to war. Increasing the number of nodes $N$ also causes the peace state to become more fragile.\footnote{We also observe that the bifurcation conditions, (\ref{eq:alpha_PW_special}) and (\ref{eq:alpha_WP_special}), are applicable to the $N=2$ case, showing that the self-tie contributions to the structural balance force dynamics noted in Sec.~\ref{sec:signed_model} result in bifurcations between peace and war in the two-node case as well.} The dependence on $L$ on the right-hand side Eq.~(\ref{eq:alpha_WP_special}) implies that the war-to-peace bifurcation sees the height of the balance force plateau, unlike the peace-to-war transition.  Since $\alpha^*_{W\rightarrow P}\rightarrow 0$ as $L\rightarrow \infty$, increasing the intensity of cooperation and conflict in the war state makes it more difficult for the system to transition back to the peace state.

\subsection{General results}
\label{sec:genresults}

We consider bifurcations in the dynamics for the general case where the dyadic bias parameters can be different for each dyad. In addition, we allow for application of a transient perturbation which can induce a transition from the peace to the war state (or more generally a state of high tie values). We state the key results in this section and show the derivations leading to them in Appendix~\ref{app:general_case}.

Our dynamical system of edges, Eq.~\ref{eq:dyn_sys_edge_ij_full}, can be written succinctly as a matrix dynamical system
\begin{align} \label{eq:dyn_sys_X_full_1}
    \frac{d\b{X}}{dt} = -\beta (\b{X} - \b{X}_D) + L \tanh \left( \frac{\alpha \b{X}^2}{L} \right) + \b{X}_T(t),
\end{align}
where $\b{X}$ is the symmetric matrix of ties $X_{ij}$, $\b{X}_D$ is the matrix of the dyadic bias parameters, $X_{Dij}$, and $\b{X}_T(t)$ is the matrix of the transient perturbations $X_{Tij}(t)$. We can analyze the stability of the  dynamics in the low-dimensional eigenvalue-eigenvector space of $\b{X}$ and derive the system's bifurcation conditions in terms of the leading eigenvalue of $\b{X}_D$.

We first state the results in the absence of an imposed perturbation $\b{X}_T(t)$. As in the special case, the peace-to-war transition is a saddle-node bifurcation. It occurs when the respective eigenvalues, $\lambda_P$ and $\lambda_U$, of the peace and unstable equilibria, $\b{X}_P$ and $\b{X}_U$,  merge. The general analysis approximates the structural balance term using $\tanh \theta \approx \theta$ for small $\theta$ as in the special case. The leading eigenvalue of the peace state is given by
\begin{align}\label{eq:lambda_P}
    \lambda_P = \frac{\beta - \sqrt{\beta^2 - 4\alpha \beta \lambda_{D1}}}{2 \alpha},
\end{align}
where $\lambda_{D1}$ is the leading eigenvalue of the dyadic bias matrix $\b{X}_D$.
The lowest eigenvalue of the unstable state is the same as the expression for $\lambda_P$, but with a positive sign in front of the square root term (see Eq.~\ref{eq:lambda_U_notrans}). Setting $\lambda_P=\lambda_U$, equivalent to the vanishing of the square root term above, yields the critical value for the structural balance sensitivity in the war-to-peace bifurcation,
\begin{align}
    \alpha^*_{P\rightarrow W} = \frac{\beta}{4 \lambda_{D1}}. \label{eq:alpha_star1}
\end{align}
After the peace-to-war bifurcation, the system converges to the war equilibrium, $\b{X}_W$, which has leading eigenvalue $\lambda_W$. Using the flat plateau approximation of the structural balance force, $L \tanh \theta \approx \pm L$ for large $|\theta|$ yields the following approximation for $\lambda_W$, 
\begin{align}
	\lambda_W = \lambda_{D1} + \frac{L N}{\beta}.
\end{align}
The bifurcation from the war state back to the peace state occurs via a different saddle-node bifurcation where $\lambda_U = \lambda_W$. The approximate condition for the war-to-peace bifurcation is then given by
\begin{align}
	\alpha^*_{W\rightarrow P} &= \frac{NL}{(\lambda_{D1} + NL/\beta)^2}.
	\label{eq:alpha_WP_gen}
\end{align}
The general expressions for the critical $\alpha$ values, (\ref{eq:alpha_star1}) and (\ref{eq:alpha_WP_gen}), are the same as in the special case except that $N\mu$ has been replaced by $\lambda_{D1}$ (the next section shows that $\lambda_{D1} = N\mu$ in the special case). Figures~\ref{fig:intro_fig} and \ref{fig:eigvals_alpha} demonstrate how the analytically computed bifurcation values match the bifurcations observed through simulation. However, the peace-to-war estimated bifurcation is more accurate than the war-to-peace estimated bifurcation.

The analysis of the system with the transient perturbation term is restricted to the case when $\b{X}_T(t)$ is applied as an impulse in which all of its elements are constant for the duration of the impulse. In this case, the impact of $\b{X}_T(t)$ can be accounted for by a modified dyadic bias matrix, $\tilde{\b{X}}_D$, that combines $\b{X}_D$ and $\b{X}_T$ via 
\begin{align} \label{eq:XDtilde}
	\tilde{\b{X}}_D(t) = \b{X}_D + \frac{\b{X}_T(t)}{\beta}.
\end{align}
The threshold condition for $\b{X}_T(t)$ to induce the system to switch from the peace to the war state is analogous to the war-to-peace bifurcation in which $\lambda_P=\lambda_U$. However, the leading eigenvalue of the dyadic bias matrix in Eq.~(\ref{eq:lambda_P}) for $\lambda_P$ as well as that for $\lambda_U$ (Eq.~\ref{eq:lambda_U_notrans}) is replaced by the leading eigenvalue of $\tilde{\b{X}}_D$, which we write as $\tilde{\lambda}_{D1}$. Accordingly, the threshold condition is the same as the peace-to-war bifurcation condition, Eq.~(\ref{eq:alpha_star1}), but with $\tilde{\lambda}_{D1}$ instead of $\lambda_{D1}$. For the imposed perturbation context, however, we keep the model parameters fixed and vary $\b{X}_T$, so it is more approprate to cast the condition in terms of the critical value of $\tilde{\lambda}_{D1}$,
\begin{align}\label{eq:lam_tilde_star1}
     \tilde{\lambda}_{D1}^* &= \frac{\beta}{4 \alpha}.
\end{align}

\section{Enhancement of instability due to polarized community structure} \label{sec:commstruc}

We now use the results of the stability analysis above to generalize the behavior seen in the simulations in Fig.~\ref{fig:eigvals_alpha} showing the greater instability of polarized two-faction community structure as compared with the no structure case, that is, its war-to-peace bifurcation occurs at a lower critical value $\alpha^*_{P\rightarrow W}$. We will show that networks whose first eigenvector is characterized by a community structure consisting of opposing factions have a lower $\alpha^*_{P\rightarrow W}$ than networks where such structure is not discernible. The lack of discernible structure may be because there is simply no tendency for preferential tie formation  built into the stochastic process that generates the dyadic bias parameters $X_{Dij}$ (also taken as the initial conditions) or, alternatively, the tendency is present but too weak to be discerned in the eigenvalue spectrum. 

We can create two-faction community structure in $\b{X}_D$ with a signed network stochastic block model which assigns a positive, negative or zero tie value between nodes with probabilities dependent on whether the nodes are drawn from the same block or different blocks \citep{morrison_community_2020}. The probability of a dyad being connected by any tie, whether positive or negative, is given by the ingroup tie density, $d_{in}$, for intra-block dyads and the outgroup tie density, $d_{out}$, for inter-block dyads. The ingroup and outgroup affinities, $p^+_{in}$ and $p^+_{out}$, for the intra and inter-block dyads respectively are the probabilities of a positive tie conditional upon a tie being present. There are two blocks of equal size with the first comprising the first $N/2$ nodes, $(1, \dots, N/2)$ and the other consisting of the next $N/2$ nodes, $(N/2+1, \dots, N)$. The generated adjacency matrices can be written as the sum of an average ``signal'' matrix $\langle \b{X}_D \rangle$ and a zero-mean random ``noise'' matrix. The signal matrix can be decomposed as the sum of two matrices formed by the outer products of its eigenvectors as follows:

\begin{align}
\label{eq:Aavg_clean}
    \langle \b{X}_D \rangle &= N \omega \b{u}_H \b{u}_H^T +   N \mu  \b{u}_C \b{u}_C^T,
\end{align}
where $\b{u}_H = \frac{1}{\sqrt{N}}[1, 1, ..., 1]^T$ and $\b{u}_C = \frac{1}{\sqrt{N}}[1,..., 1, -1,..., -1]^T$ are orthonormal vectors (the first negative component of $\b{u}_C$ is at $N/2+1$).  The first term generates an $N \times N$ matrix all of whose elements are equal to $\omega$, the expected average of all the tie values in $\langle \b{X}_D \rangle$. The second term generates a matrix where the block diagonal ties are given by $+\mu$ and the off-diagonal blocks are $-\mu$. $\b{u}_H$ and $\b{u}_C$ are the signal eigenvectors. As $\b{u}_H$ generates a uniform matrix, it is called the ``homogeneous'' eigenvector, whereas $\b{u}_C$ is the ``contrast'' eigenvector as it generates the difference between intra-block and inter-block behavior. The noise matrix then accounts for the fact that, probabilistically, some ties will be absent or of opposite sign than expected for their block. 

The contrast parameter $\mu$ controls the level of polarization in the network. The block structure resulting from $\mu>0$ could reflect the operation of a homophily process, which results in preferential tie formation due for instance to similar ideology, religion, or ethnicity (the $\mu<0$ case is not relevant here). The relative values of $\omega$ and $\mu$ determine which kind of signal structure dominates: homogeneous if $\omega  > \mu$ and contrast if $\mu > \omega$.  The eigenvalues of $\b{u}_H$ and $\b{u}_C$ are given by, respectively, $\lambda_H=N\omega$ and $\lambda_C=N\mu$. Consequently, we can also use the relative values of $\lambda_H$ and $\lambda_C$ to assess whether homogeneous or contrast structure dominates the signal. Note that the special case of equal and opposite factions above in which all intra-faction dyadic biases are $+\mu$ and all inter-faction ones are $-\mu$ can be completely constructed by the contrast eigenvector so that $\b{X}_D = N \mu \b{u}_C \b{u}_C^T$. The first (and only) eigenvalue of $\b{X}_D$ is then $N\mu$. This connects the special case result for the critical point $\alpha^*_{P\rightarrow W} = \beta /(4 N \mu)$ (Eq.~\ref{eq:alpha_PW_special}) with the general result $\alpha^*_{P\rightarrow W} = \beta /(4 \lambda_{D1})$ (Eq.~\ref{eq:alpha_PW_general}).

If a signal eigenvalue is sufficiently large, then it will appear outside of the dense band of eigenvalues due to the noise. For example, in Fig.~\ref{fig:homog_fact}(a), the contrast eigenvalue starts out as the leading eigenvalue at the bottom of the plot,  becomes the second eigenvector after the signal crossing at $p^+_{out}=0.5$, and then merges into the noise band around $p^+_{out}=0.8$. This point corresponds to a phase transition for large $N$, known as the community detectability transition \citep{nadakuditi_graph_2012, morrison_community_2020}. Past the detectability transition, even though the stochastic model still favors ingroup positive ties, $p^+_{in}>p^+_{out}$, this community structure cannot be identified given only the observed network.

With this framework relating community structure to the adjacency matrix spectrum, we can address how community structure in the network of dyadic bias parameters affects the peace-to-war bifurcation threshold. The first eigenvector of  $\b{X}_D$ contains community structure if it has components of opposite sign. Equation~(\ref{eq:alpha_PW_general}) states that $\alpha^*_{P\rightarrow W}$ is inversely proportional to the first eigenvalue, $\lambda_{D1}$, of  $\b{X}_D$.  If $\lambda_{D1}$ is then also outside the noise band,  $\alpha^*_{P\rightarrow W}$ will be lower as compared to when $\b{X}_D$ is generated from the noise only. Consequently, the dynamical system in this case will be more prone to become unstable with respect to changes in system parameters that increase the balance sensitivity $\alpha$. However, the operation of a homophily process will not enhance instability if it is weak enough so that its spectral signature lies below the noise band.

Although the above argument is based on two equal-sized blocks, stochastic block models with multiple, unequal blocks also display a spectral signature consisting of a noise band and outlying eigenvalues.\footnote{The relationship between the phase transition in spectral structure and community detectability, however, only holds for equal size blocks and breaks down for unequal ones \citep{ZhaMooNew2016}.} As long as the first adjacency matrix eigenvector contains community structure and its eigenvalue is discernible outside the noise band, then the system will be more sensitive to the peace-to-war bifurcation than when the structure is absent or submerged in the noise band. Therefore, instability will be enhanced whenever community structure is present in the first eigenvalue of  $\b{X}_D$, regardless of whether the blocks in the generating model are equal in size or how many blocks there are.

Random matrix theory provides tools to analyze the noise matrix and so quantitatively assess the question of whether the first eigenvalue lies outside the band \citep{PottersBouchaud2021}. Expressions for the eigenvalue spectrum have been derived such as the Wigner semicircle law. In particular, the mean value of the band edge can be calculated as a function of the noise variance and the number of nodes, and so can the expected deviation from the mean. Consequently, the first eigenvalue can be taken as signal if it is sufficiently far outside the band edge expected from noise only.

\section{Destabilizing directions of finite perturbations}

Complex systems often exhibit nonlinear dynamics that have multiple stable states between which the system can transition \citep{strogatz_nonlinear_2016, Roberts2015}. Random perturbations can induce transitions between states but it is also possible to investigate the effect of particular perturbations. Methods integrating bifurcation and control theories have been used to understand how targeted perturbations can be employed for control purposes in nonlinear dynamical networks, such as the hypothesized use of transient signals to shift between stable states in nematode locomotion or jumping between memory states in artificial neural networks \citep{ morrison_nonlinear_2021, morrison_nonlinear_2021-1}. However, these methods can also be applied to the identification of deleterious perturbations, such as those that might trigger a major war. Identifying such perturbations may ultimately help in preventing them.

We now turn to the situation where our dynamical system, (\ref{eq:dyn_sys_X_full_1}), can be shifted from one equilibrium to another, not by parameter changes, but by finite perturbations as contained in the matrix $\b{X}_T$.  Our goal here is to discover which patterns of dyads, if perturbed, are most dangerous in terms of destabilizing the peace state. We investigate this by determining what perturbations, $\b{X}_T$, of a given size, $||\b{X}_T||=c$, move the system closest to the critical bifurcation threshold; these are the perturbation directions that are most dangerous as they are most likely to cause destabilization if they are too large. We measure how close the system is to a bifurcation by measuring how close the leading eigenvalue is to the critical bifurcation eigenvalue.  First, we develop the mathematical framework for identifying these patterns according to different optimization conditions. We then illustrate how this framework can be applied using the network of five great power just before the onset of World War I. 

Destabilization of the peace state can be achieved through a change in model parameters as occurred in Fig.~\ref{fig:intro_fig}(b) when the balance sensitivity was increased from 0.05 to 0.1; the peace equilbrium disappeared and the system shifted to the war state, the only equilibrium remaining. However, $\alpha=0.05$ resides in the bistable regime where both the peace and war equilibria are stable, as seen in Fig.~\ref{fig:intro_fig}(c). Consequently, an alternative mode of destabilizing the peace state in the bistable regime is via a perturbation strong enough to kick the system into the war state.  These perturbations may be localized to particular nodes, for example, a military crisis between two states.  They are also assumed to be transient in duration so that, absent an escalation to system-wide war, the perturbed dyads will return to their equilibrium tie values.

In our stability analysis results for the general case (Sec.~\ref{sec:genresults}),  perturbations in $\b{X}_T$ were incorporated into a modified bias matrix, $\tilde{\b{X}}_D$, with leading eigenvalue, $\tilde{\lambda}_{D1}$. To shift from the peace to the war state in the bistable regime then requires that $\tilde{\lambda}_{D1}$ equal or exceed the eigenvalue of the unstable equilibrium, $\lambda_{U}$, with the critical eigenvalue denoted by $\tilde{\lambda}^*_{D1}$. The effect of the perturbation can therefore be seen as equivalent to a temporary change in the parameters of the system so that a saddle-node bifurcation is induced. This condition is given by  Eq.~(\ref{eq:lam_tilde_star1}), $\tilde{\lambda}^*_{D1}=\beta/4\alpha$, for the peace-to-war bifurcation. 
We analyze the types of perturbations, $\b{X}_T$, that can most readily destabilize a system for a given $\b{X}_D$.
We first consider a special case for the perturbation matrix where $\b{X}_T$ can be analytically calculated. We then present  computational schemes to identify other destabilizing perturbations including those that may produce final states other than the war state that results from the peace-to-war bifurcation.

\subsection{Minimum energy destabilization using dyadic bias leading eigenvector}
\label{sec:minenergy}

The perturbation to the peace state that requires the least amount of energy, $||\b{X}_T||$, to destabilize is along the direction of the leading eigenvector of $\b{X}_D$, denoted by $\mathbf{s}_{D1}$. The resulting perturbation matrix can be written as $\b{X}_T =\sigma  \mathbf{s}_{D1} \mathbf{s}_{D1}^T$.  We solve for the mimimum value of $\sigma$ that will destabilize the system ($||\b{X}_T||=\sigma$). Destabilization requires that the leading eigenvalue of the  modified dyadic bias matrix, $\tilde{\lambda}_{D1}$, is at least as large as the critical eigenvalue so that 
\begin{align}
	\tilde{\lambda}^*_{D1} & \leq \tilde{\lambda}_{D1}.
\end{align}
Making use of Eq.~(\ref{eq:lam_tilde_star1}) and the definition of $\tilde{\b{X}}_{D}$, (\ref{eq:XDtilde}), then gives
\begin{align}
 	\frac{\beta}{4 \alpha} & \leq \mathbf{\lambda_1} \left(\b{X}_D + \frac{\b{X}_T}{\beta} \right)  \\
    & \leq \mathbf{\lambda_1} \left(\b{X}_D + \frac{\sigma}{\beta} \mathbf{s}_{D1} \mathbf{s}_{D1}^T \right)  \\
	& \leq \lambda_{D1} + \frac{\sigma}{\beta}
\end{align}
where the notation $\mathbf{\lambda_1}(\cdot)$ refers to the first eigenvalue of its argument and we have used $\b{X}_D \b{s}_{D1} = \lambda_{D1} \b{s}_{D1}$ and $\mathbf{s}_{D1}^T \mathbf{s}_{D1} = 1$ to show that $\b{s}_{D1}$ is also an eigenvector of $\tilde{\b{X}}_D$. Solving for $\sigma$ we find yield the condition required for destabilization,
\begin{align} \label{eq:optimal_sigma}
	\sigma \geq \beta (\frac{\beta}{4 \alpha} - \lambda_{D1}).
\end{align}
It is also useful to write this condition as 
\begin{align} \label{eq:optimal_sigma_alt}
	\sigma \geq \beta (\tilde{\lambda}^*_{D1} - \lambda_{D1}).
\end{align}

\subsection{Destabilizing directions for  perturbations}

Although $\b{X}_T = \sigma \b{s}_{D1} \b{s}_{D1}^T$ is the perturbation that destabilizes the system with the minimum amount of energy (minimum $||\b{X}_T||$), there are other, slightly higher energy perturbations that can also destabilize the system. We can search for perturbation directions within the space of perturbations $||\b{X}_T||=\sigma$ that come closest to destabilizing the system. These are the perturbation directions that are most likely to destabilize the system since they require less energy to destabilize the system than other perturbations.
We use the following optimization scheme to find perturbation directions $\b{X}_T$ that are closest to destabilizing the system for a given energy level $||\b{X}_T|| = \sigma$: 

\begin{align}
        \text{Energy minimizing scheme:} \quad &\argmin_{\b{X}_T \in A}\left[  ||\tilde{\lambda}_{D1}^* - \tilde{\lambda}_{D1}|| \right] \label{eq:control_min_energy}
\end{align}
where $A = \{\b{X}: ||\b{X}|| = \sigma\}$, $\tilde{\lambda}_{D1}$ is the leading eigenvalue of $\tilde{\b{X}}_D$, and $\tilde{\lambda}_{D1}^*$ is the stability threshold (Eq.~\ref{eq:lam_tilde_star1}). Energy minimizing perturbations result in minimal changes, on average, in tie strengths, whether positive or negative, since $||\b{X}_T||$ has been minimized. Eq.~(\ref{eq:control_min_energy}) brings the system close to destabilization by perturbing the system in a direction close to the leading mode.

We now present two alternative destabilization schemes. The low-rank war state structure that appears after destabilizing the peace state will not necessarily contain two equally sized factions. The leading eigenvector could instead contain all nodes or mostly all nodes of the same sign. The interpretation of destabilization in this case would be instead the formation of a tightly-knit alliance structure where, due to structural balance dynamics, nodes are compelled to form positive ties with other nodes due to mutual allies. We can use the ``eigenvector polarization'' measure of the leading eigenvector, $\phi_1$, to determine the extent to which the leading eigenvector contains community structure or a lack of community structure close to homogeneity \citep{morrison_contribution_2021}. The eigenvector polarization $\phi_i$ of eigenvector $\b{s}_i$ of the adjacency matrix $\b{X}$ is $\phi_i = \b{s}_i^T \b{M} \b{s}_i$, where $\b{M}$ is the signed modularity matrix \citep{traag_partitioning_2019, newman_finding_2006}. 
The modularity matrix clusters nodes into communities.\footnote{$\b{M}=\b{X}-\frac{\b{k}\b{k}^T}{2m}$, where $\b{k}$ is the vector of node degrees and $m$ is the total number of edges.} If $\phi_1$ is the largest eigenvector polarization, then the network structure is dominated by polarization between two hostile factions.

We use $\phi_1$ to define the following two schemes, which seek to: (1) bring the system close to destabilization with a widespread two-faction conflict (Eq.~\ref{eq:control_fact}), or, alternatively, (2) bring the system close to destabilization with a widespread alliance bloc (Eq.~\ref{eq:control_homog}).
\begin{align}
    \text{Polarizing:} \quad &\argmin_{\b{X}_T^F \in A}\left[  ||\tilde{\lambda}_{D1}^* - \tilde{\lambda}_{D1}|| + \epsilon \frac{1}{\phi_1} \right] \label{eq:control_fact}\\
    \text{Harmonizing:} \quad &\argmin_{\b{X}_T^H \in A}\left[  ||\tilde{\lambda}_{D1}^* - \tilde{\lambda}_{D1}||  + \epsilon \phi_1 \right] \label{eq:control_homog}
\end{align}
Here, $\phi_1$ is the eigenvector polarization measure of $\tilde{\b{X}}_D$, which changes as the perturbation $\b{X}_T$ is varied, and  $\epsilon$ is its weighting.
 Forcing large changes in ties, whether increasing positive relations or increasing hostility requires more ``energy" (a larger $||\b{X}_T||$) than smaller perturbations in relations between countries. Because the leading eigenvector of the adjacency matrix may not contain factions, energy minimizing perturbations may not result in factions. The polarizing perturbation includes a faction-promoting term $\frac{1}{\phi_1}$, weighted by $\epsilon$, that is minimized when $\phi_1$ is large. This promotes a solution where the leading eigenvector of the system contains factions. 
 The extent to which factions are promoted is determined by the strength of $\epsilon$ (small enough values for $\epsilon$ may not produce solutions that contain factions). The harmonizing perturbation, in contrast, deincentivizes factions in the leading eigenvector by including the faction-suppressing term $\phi_1$. The homogeneous perturbation is minimized when $\phi_1$ is small which promotes a solution where the leading eigenvector of the system does not contain factions. These perturbation schemes assume that $\b{X}_T$ can be made large enough to result in a transition from the peace state to an equilibrium with high conflict and cooperation tie values. If the maximum structural balance force strength, $L$, is sufficiently large, then the final state of the system will be determined by the first eigenvector of $\tilde{\b{X}}_D$ (see Eq.~\ref{eq:high_state}). Therefore, as $\b{X}_T$ can cause the leading eigenvector of $\tilde{\b{X}}_D$ to be very different from that of $\b{X}_D$, the high tie value state produced by the perturbation may contain very different factions than those in the war state arising from the peace-to-war bifurcation.

\section{Application to onset of World War I} \label{sec:WWI}

\subsection{War onset due to triggering perturbations}

We use the maximally destabilizing directions formulas, Eqs.~(\ref{eq:control_min_energy}), (\ref{eq:control_fact}), and (\ref{eq:control_homog}), to find destabilizing directions for perturbations $\b{X}_T$ for the network of great powers leading up to the First World War under the dynamics of Eq.~(\ref{eq:dyn_sys_X_full_1}). World War I is considered a classic example of a systemic war that emerged from the complex dynamics of five European great powers, the United Kingdom, France, Germany, Austria-Hungary, and Russia. We will compare our computationally derived destabilizing perturbation directions to ties that have been historically identified as critical to the outbreak of the Great War.

We initialize the system using empirical data as follows. $\b{X}_D$ is set to the edge values in the year 1913 using the militarized interstate disputes (MIDs), alliances, and rivalries data from the Correlates of War dataset and the Handbook of International Rivalries \citep{palmer_mid4_2015,gibler_international_2009,singer_formal_1966,thompson_handbook_2011}. Alliances contribute positive weights, MIDs contribute negative weights for states on opposite sides of a dispute and positive weights for states on the same side of a dispute, and rivalries contribute negative weights. Weights between states are added and scaled to range from -2 (rivals) to +2 (allies).  The dyadic force strength is set to $\beta = 1$ and the balance sensitivity to $\alpha = 0.03$.

\begin{figure}[th!]
    \centering
    \includegraphics[width=0.85\linewidth]{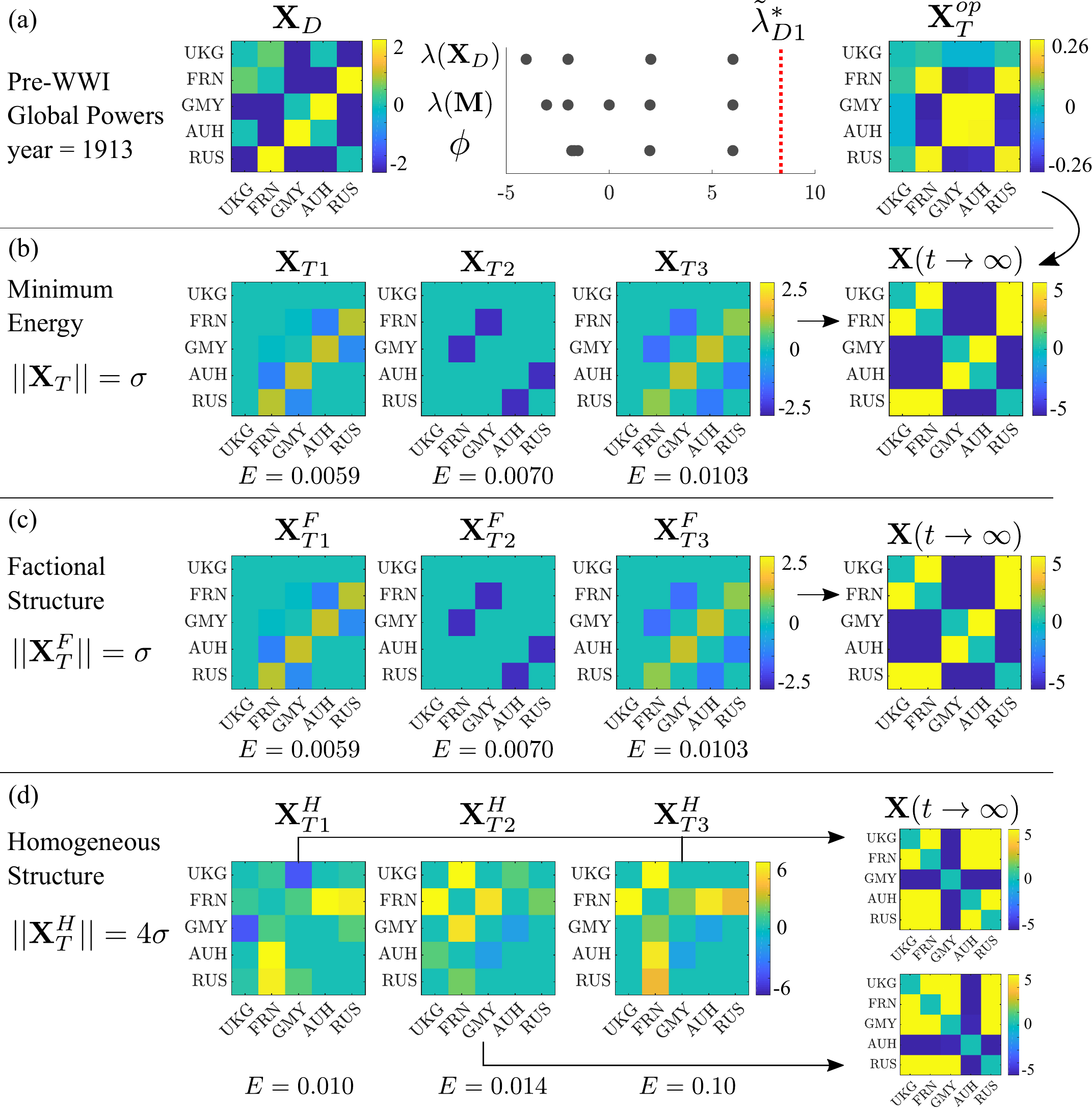}
    \caption{Equilibrium pre-WWI network under edge dynamics with destabilizing perturbation directions $\b{X}_T$. (a) $\b{X}_D$ is set to the edge weights of the great powers in 1913, immediately preceding WWI. For $\alpha = 0.03$ and $\beta=1$, the bifurcation threshold is $\tilde{\lambda}_{D1}^* = 8.33$. The leading eigenvalue of $\b{X}_D$ is $\lambda_{D1}=6.01$ resulting in $\sigma = 2.32$. Minimum energy perturbation $\b{X}_T^{op}$. (b) Perturbation directions requiring minimal energy (Eq.~\ref{eq:control_min_energy}). Perturbations applied in this direction result in two factions in the leading eigenvector, and therefore two factions in the high-conflict state, $\b{X}(t \rightarrow \infty)$. (c) Maximally destabilizing directions that include factional structure (Eq.~\ref{eq:control_fact}), result in two factions in the leading eigenvector. (d) Maximally destabilizing directions that have homogenous structure (Eq.~\ref{eq:control_homog}), results in a leading eigenvector with an almost homogenous struction. Error minimized, $E = \tilde{\lambda}_{D1}^* - \tilde{\lambda}_{D1}$. Eq.~\ref{eq:high_state} approximates equilibrium war states.}
    \label{fig:ww1_control_sigs}
\end{figure}

The left panel of Fig.~\ref{fig:ww1_control_sigs}(a) shows the dyadic bias matrix, $\b{X}_D$ while the middle panel plots the eigenvalues of $\b{X}_D$, the eigenvalues of the modularity matrix $\b{M}$ associated with $\b{X}_D$, and the associated eigenvector polarization $\phi$  of all the eigenvectors. The leading eigenvalue of  $\b{X}_D$ has value $\lambda_{D1}= 6.01$, which is observed to be well separated from the remainder of the spectrum as is the first eigenvalue of the modularity matrix. The fact that $\phi_1$ is the largest eigenvector polarization signifies that $\b{X}_D$'s first eigenvector $\b{s}_{D1}$ contains contrast-like community structure consisting of hostile factions \citep{morrison_contribution_2021}. Also displayed is the critical eigenvalue of the modified dyadic bias matrix for destabilization, calculated from Eq.~(\ref{eq:lam_tilde_star1}) to be  $\tilde{\lambda}_{D1}^* = 8.33$. The optimum perturbation $\b{X}_T^{op}$ that enables destabilization with the minimum energy is displayed in the right panel of Fig.~\ref{fig:ww1_control_sigs}(a). It is proportional to the matrix formed by the first eigenvector of $\b{X}_D$ and is given by $\b{X}_T^{op} =\sigma  \mathbf{s}_{D1} \mathbf{s}_{D1}^T$ as explained in Sec.~\ref{sec:minenergy}. The minimum energy is calculated from Eq.~(\ref{eq:optimal_sigma_alt}), which yields $\sigma = 2.32$. 

Substantively, the optimal perturbation matrix, $\b{X}_T^{op}$, is marked by two tight factions -- Germany (GMY) and Austria-Hungary (AUH) versus France (FRN) and Russia (RUS); the United Kingdom (UKG) is loosely aligned with the Franco-Russian bloc. This structure is consistent with the antagonisms and alliances that were most entwined in the several years before the war \citep{taylor_struggle_1954, kagan_origins_1995}  Russia and Austria-Hungary competed for influence in the Balkans while France and Germany had severe territorial disputes and a history of animosity. France and Russia were allies as were Germany and Austria-Hungary. It is therefore reasonable that a crisis which activated tensions between the members of these alliances would be most efficient in triggering a war. In contrast, although the United Kingdom was aligned with France and Russia as a member of the Triple Entente, it was reluctant to get involved in Continental conflicts and its participation in the event of a war was uncertain.

The rightmost panel of Fig.~\ref{fig:ww1_control_sigs}(b) shows the equilibrium state, $\b{X}(t\rightarrow \infty)$, to which our dynamical model evolves the system as a result of a transient perturbation in the direction of $\b{X}_T^{op}$. The war state reached by the system matches the two opposing sides present during WWI with the United Kingdom, France, and Russia in one faction and Germany and Austria-Hungary in the other.

We solve for alternative, sparse perturbation directions using the optimization functions delineated above, setting $\sigma = 2.32$, the energy level for the optimal solution. We compute the error as the amount to which the leading eigenvalue deviates from the stability threshold $E = \tilde{\lambda}_{D1}^* - \tilde{\lambda}_{D1}$. The error tells us how close the system with a perturbation $||\b{X}_T||=\sigma$ in the given direction is to destabilizing. Smaller values mean that the system is close to a bifurcation.

Figure~\ref{fig:ww1_control_sigs}(b) shows three optimal sparse perturbation directions, $\b{X}_T$, computed using Eq.~\ref{eq:control_min_energy} from 100,000 randomly generated perturbations. These sparse perturbations have a substantial component in the direction of $\b{X}_T^{op}$. All three involve only the Continental powers. The middle one, $\b{X}_{T2}$, depicts flare-ups in the two core rival dyads, France vs. Germany and Russia vs. Austria-Hungary. It was the assassination of Austrian Archduke Franz Ferdinand by a Serbian nationalist that set off a crisis between Austria-Hungary and Russia, Serbia's ally, which ultimately led to the war. The right-hand perturbation, $\b{X}_{T3}$, reflects increased cooperation between Germany and Austria-Hungary and between France and Russia as well as escalating tensions within the Austro-Hungarian--Russian and German--French rivalries. During the crisis, both Germany and France promised to come to the aid of their respective allies in the event of war whereas Britain, absent from this perturbation, was much more ambiguous about its intentions \citep{Kissinger1994}. The lowest energy one, $\b{X}_{T1}$, represents an alternative scenario in which the conflicting dyads are France vs. Austria-Hungary and Russia vs. Germany with, again, support between allies. As with the optimal perturbation above, all three perturbations when evolved by the dynamical model converge to the same two factions as present during WWI.

Figure~\ref{fig:ww1_control_sigs}(c) shows sparse perturbations computed using Eq.~\ref{eq:control_fact}. As the leading eigenvector is already factional, the factional solutions are similar to the overall minimum energy solutions in Fig.~\ref{fig:ww1_control_sigs}(b) and lead to the two factions determined by the system's initial community structure. These sparse perturbations have a substantial component in the direction of $\b{X}_T^{op}$. The solutions in Figure~\ref{fig:ww1_control_sigs}(b,c) show that the United Kingdom's ties with Continental Europe do not appear as significant in triggering conflict. This aligns with historical accounts that the United Kingdom was reluctant to get involved in Continental conflicts \citep{taylor_struggle_1954, kagan_origins_1995}.

Figure~\ref{fig:ww1_control_sigs}(d) shows the homogenous perturbations $\b{X}_T^H$ computed using Eq.~\ref{eq:control_homog}.  The factional structure is much more dominant than other structures in the network resulting in a larger perturbation required to obtain a homogeneous final state, $||\b{X}_T^H|| = 4\sigma$. Ideally, the homogenous perturbations would result in a homogeneous structure among all nodes in the final equilibrium state. With a low level of perturbation energy, however, the system only manages to positively coalesce four of the five countries in a positive alliance bloc.
Perturbations $\b{X}_{P1}^H$ and $\b{X}_{P3}^H$ lead to a homogenous coalition against Germany while $\b{X}_{P2}^H$ leads to a homogenous coalition against Austria-Hungary.

\subsection{War onset due to changes in dyadic bias community structure}

\begin{figure}[t]
    \centering
    \includegraphics[width=0.75\linewidth]{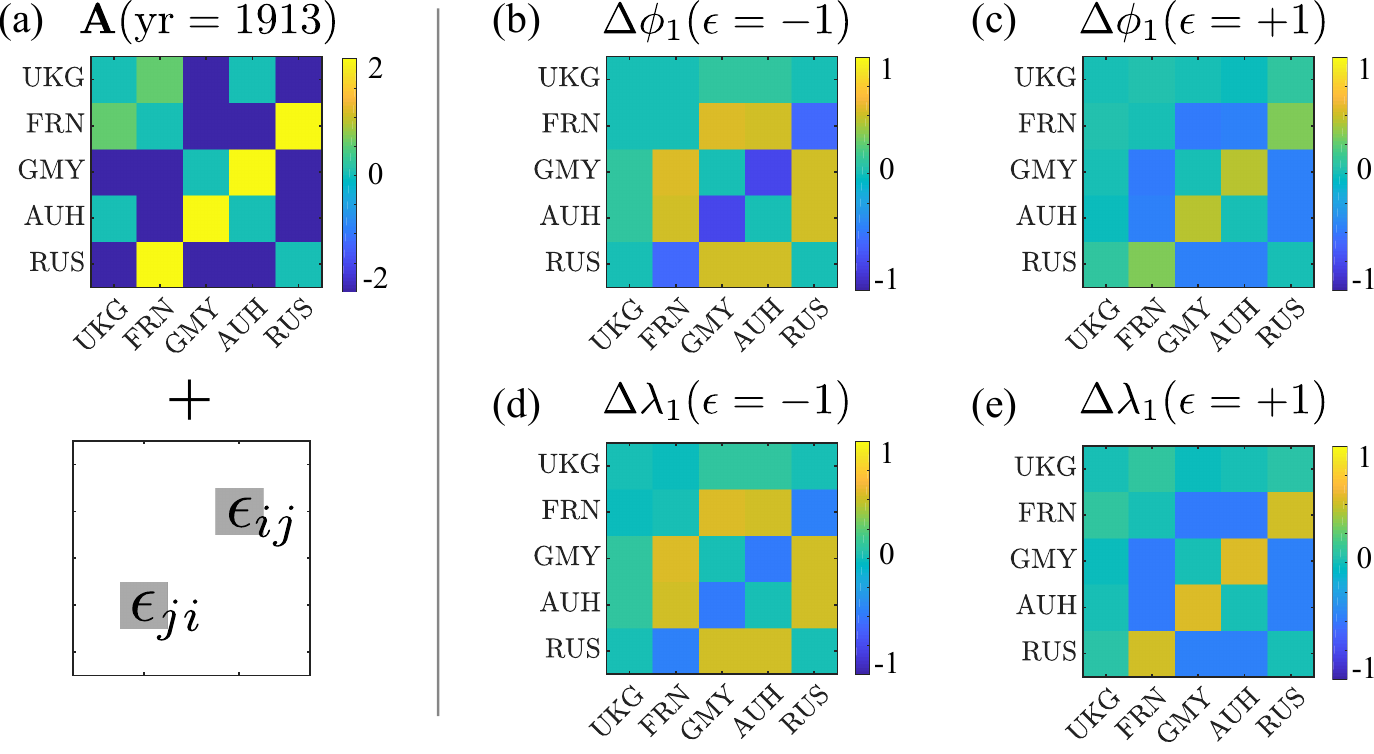}
    \caption{(a) Great powers network in 1913. (b) Effects on $\phi_1$ for a $\epsilon = -1$ edge perturbation. (c) Effect on $\phi_1$ for a $\epsilon = +1$ perturbation. (d) Effect on $\lambda_1$ for a $\epsilon = -1$ perturbation. (e) Effect on $\lambda_1$ for a $\epsilon = +1$ perturbation.}
    \label{fig:perturb_phi}
\end{figure}

We now take a different approach and evaluate which individual ties are most critical in increasing the leading eigenvalue of the dyadic bias matrix, $\b{X}_D$, and polarization in the network in the year 1913, preceding the outbreak of WWI.  Given that increasing the first eigenvalue and polarization make the system more susceptible to war, this analysis can identify which dyads most destabilize the system (or stabilize it if the first eigenvalue decreases). In Figure~\ref{fig:perturb_phi} we systematically shift the ties of $\b{A}$ by $\delta \pm 1$ and observe the change this produces in the leading eigenvalue $\lambda_1$ and polarization $\phi$, of $\b{A}$.
Figure~\ref{fig:perturb_phi}(b-e) shows that the tie between France and Germany as well as the tie between Germany and Austria-Hungary, were the most crucial in increasing community structure and polarization in the network. The ties between Great Britain and the Continential countries appear to be the least important, implying that while Great Britain formed an important part of the network, changes in ties with this state least impact the stability of the system.
Other significant ties include FRN-AUH, FRN-RUS, GMY-RUS, and AUH-RUS. This finding is to be expected as the German alliance with Austria-Hungary and their disputes with Russia as well as the alliance between Russia and France were all factors that proved integral to the development of World War I \citep{kagan_origins_1995, taylor_struggle_1954}.

\section{Critical slowing down near bifurcations} \label{sec:slowing_down}

The destabilization analyses in the preceding section required knowledge of the dynamical system governing state dynamics and therefore the location of the bifurcation point could be calculated. However, the true model governing the dynamics of real world complex systems --- physical, biological, and social --- is often unknown in the sense that precise and accurate functional forms and parameter values are not available. This makes it difficult to empirically determine the distance in parameter space from a bifurcation. Yet, the fact that the behavior in the vicinity of a bifurcation is universal for all systems that undergo that type of bifurcation can help with this task. The theory of critical transitions seeks to develop early warning signals of impending bifurcations based on this universality \citep{kuehn_mathematical_2011,  scheffer_anticipating_2012}. One common signature is critical slowing down, in which the transients due to perturbations take longer to die down as the bifurcation is neared. Critical slowing down has been empirically observed in transitions in the climate,  ecosystems, and neurons \citep{Dakosetal2008, BurSujPav2021, meisel_critical_2015}. 

Saddle-node bifurcations exhibit critical slowing down and so it should occur in our model dynamics and could be used to infer distance to the bifurcation in a real system in the absence of exact governing equations. Meisel \textit{et al.} (\citeyear{meisel_critical_2015}) found that for saddle-node bifurcations occurring in neural dynamics the recovery rate $r=\frac{1}{\tau}$, where $\tau$ is the decay time, scales with the square root of the distance from the bifurcation, $r \propto \sqrt{d}$, a scaling law that can be derived from the saddle-node normal form. Accordingly, we investigate whether our model follows this same scaling law. 

\begin{figure}[t]
    \centering
    \includegraphics[width=0.85\linewidth]{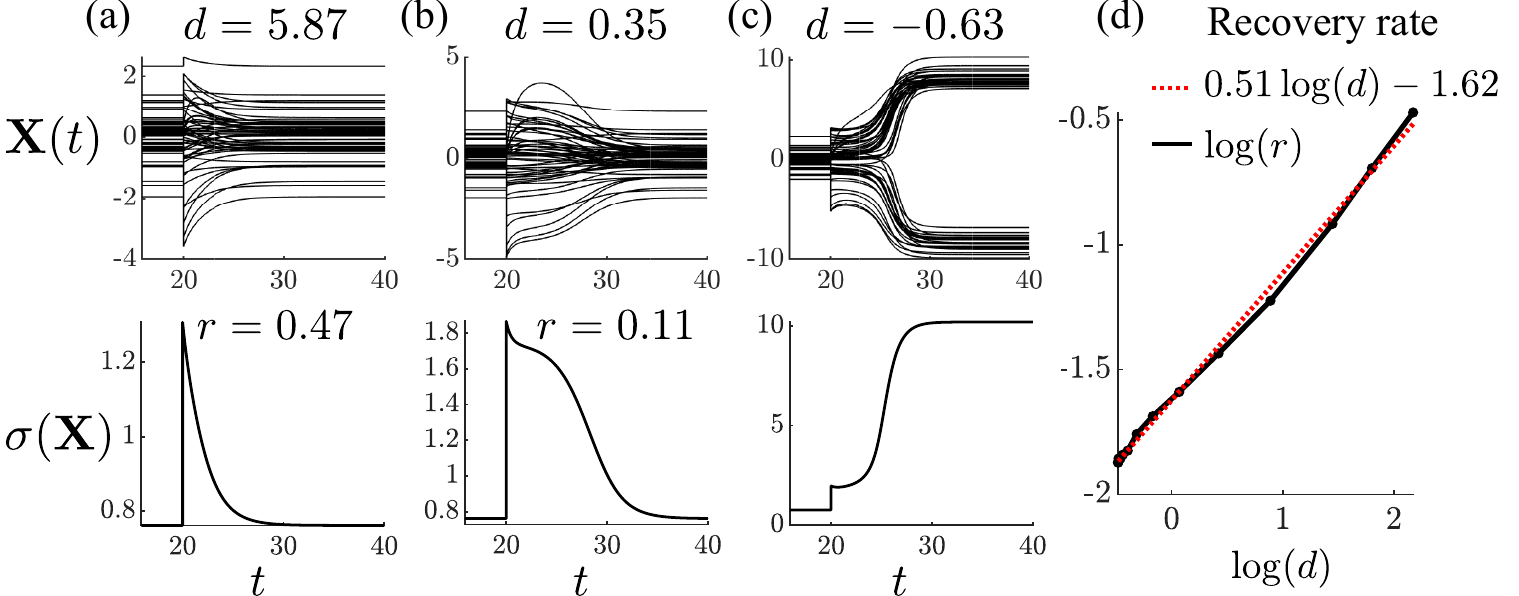}
    \caption{Critical slowing down near bifurcation. Recovery rate $r$ is a function of the distance to the bifurcation in the eigenvalue space $d = \lambda_1 - \lambda_U$. (a) Perturbation to the system far from the unstable fixed point, $d = 5.87$, results in a fast recovery rate, $r = 0.47$. (b) Perturbation close to the unstable fixed point, $d = 0.35$, results in a slow recovery rate, $r=0.11$. (c) Perturbation past the unstable fixed point $d = -0.63$ destabilizes the peaceful state. (d) The log of the recover rate ($r$) scales linearly with the log of the distance to the bifurcation  ($d$), $r \approx 0.2 \sqrt{d}$.}
    \label{fig:critical_slow}
\end{figure}

We measure the standard deviation of the edges after increasingly large perturbations to Eq.~(\ref{eq:dyn_sys_X_full_1}) in the peace state and compare the distance to the bifurcation, $d = \lambda_1-\lambda_U$, with the recovery rate $r$ of the standard deviation. All perturbations are in the direction of the outer product of the leading eigenvector, $\b{v}\b{v}^T$. With perturbations that keep the system far from the bifurcation, the network activity decays quickly back to equilibrium (Fig.~\ref{fig:critical_slow}(a)) while with perturbations that move the system close to the bifurcation, the network activity exhibits a delay in its decay back to equilibrium (Fig.~\ref{fig:critical_slow}(b)). The system moves to the war state for perturbations past the critical value (Fig.~\ref{fig:critical_slow}(c)). Figure~\ref{fig:critical_slow}(d) shows that $\log(r)$ scales linearly with $\log(d)$. A linear fit to the data finds that $\log(r) = 0.51 \log(d)-1.62$. This means that the recovery rate and distance to the bifurcation are related by $r \approx 0.2 d^{\, 0.51}$ in good agreement with the square root scaling  predicted for saddle-node bifurcations.

The scaling law may be used to estimate the distance to a bifurcation based on empirical measurements of the recovery rate. We investigated using recovery rates to assess bifurcation proximity for the four decades prior to World War I for the network of the five great powers. We used the alliance and militarized interstate disputes (MIDs) data from the Correlates of War dataset and rivalries from the Handbook of International Rivalries as in Sec.~\ref{sec:WWI}. The tie values in the equilibrium state of the system were taken to be the alliances and rivalries which endure on a relatively long timescale in comparison with the MIDs, which emerge and decay rapidly. Hence, the MIDs were taken to be the perturbations. The recovery rate in the standard deviation was measured after each MID, which was then used to estimate the distance to the bifurcation. The recovery rates showed significant variance, however, and they did not display a consistent decrease as the war grew closer, preventing consistent estimates of the distance to the bifurcation. Using the full network of European states would provide more complete data in which to test critical slowing down.

\section{Discussion} \label{sec:discussion}

This section discusses connections between our dynamical systems model of signed networks and risk factors and signatures associated with conflict escalation as identified in the literature on international relations. These connections include the role of nonlinear interactions,  community structure and alliance polarization, network size, system variables that could play the role of a bifurcation parameter,  perturbations and the catalysts of systemic war, and empirical implications of war as bifurcation.

Systemic wars, those involving a breakdown of the international system that entrains most if not all of the great powers in a region have been classified as either ``wars of mobilization''  or ``structural wars'' \citep{midlarsky_hierarchical_1986}. Wars of mobilization arise from the overtly hegemonic and expansionist aspirations of a revolutionary or revisionist state, such as Napoleonic France or Nazi Germany.  Structural wars are instead attributed to a complex interaction of factors including rival alliances, security dilemmas, and triggering crises. Examples include the Peloponnesian Wars, the Thirty Years' War, and World War I \citep{kagan_origins_1995, kagan_peloponnesian_2004, midlarsky_hierarchical_1986}. Structural wars   anchor our perspective here as their onset is characterized by instability rather than openly belligerent intent.

Nonlinear interactions between tightly coupled relationships of rivalry and alliance have been theorized as characterizing the lead up to systemic wars \citep{Lebow2000, thompson_streetcar_2003}. In our model, these nonlinear interactions appear in the structural balance term which contains the products of the values of adjacent ties. However, unlike the pure structural balance dynamics of Eq.~(\ref{eq:kulakowski}), the presence of the direct dyadic force competes with the balance force, seeking to pull a relationship back to its bias value. This competition and the nonlinear nature of the balance force give rise to a bifurcation between the low and high tie strength equilibria, and so the dynamics of the model itself provides an intrinsic distinction between peace and war as two qualitatively different states. Transitions between peace and war, which can go in either direction, then correspond to changes in the equilibrium state of the system. Consequently, there is no need to set an arbitrary threshold between peace and war as would be the case in a linear model where the equilibrium would change smoothly, not sharply, as the relative strength of the dyadic and balance forces is changed. 

\subsection{Polarized community structure and network size} \label{sec:polarization}

Factors thought to influence the stability of the international system can be investigated in  our model via their effect in facilitating shifts between the peace and war states. The emergence of two rival alliances, sometimes referred  to as bipolarization, is a factor commonly believed to increase the chance of systemic war, as occurred prior to the Peloponnesian War and World War I \citep{Wayman1984, thompson_streetcar_2003, kagan_peloponnesian_2004}.  Our finding that polarized networks, those whose dominant structure consists of two opposed factions, are most prone to instability provides a dynamical basis for the dangers of bipolarization. This result stems from the fact that the critical value of the structural balance sensitivity at which the peace-to-war bifurcation occurs is inversely proportional the first eigenvalue of the matrix of dyadic biases (Eq.~(\ref{eq:alpha_PW_general})). The first eigenvalue increases as polarization becomes more dominant so that a lower balance sensitivity is needed to transition to the war state as seen in Fig.~\ref{fig:eigvals_alpha}. Polarization is the dominant structure when the first eigenvector has both positive and negative components and its eigenvalue lies outside the main band formed by the noise eigenvalues (Sec.~\ref{sec:commstruc}). The polarization metric, which measures the alignment of the modularity matrix and the first eigenvector, can be used to distinguish between when the first eigevector is polarized and when its components are of uniform sign, corresponding to homogeneous structure \citep{morrison_contribution_2021}.

The instability-enhancing effect of polarized community structure arises naturally from the interplay of the dyadic and balance forces in our model. The more polarized the dyadic bias matrix is, the more balanced it is. At the limit, the matrix formed by the contrast eigenvector is perfectly balanced. Given that the tie values of the system in the peace equilibrium lie near their dyadic bias values, a polarized peace state will experience a greater balance force than a non-polarized state; the argument of the hyperbolic tangent function in Eq.~(\ref{eq:dyn_sys_edge_ij_full}) for the polarized state will have a larger magnitude because its greater balance implies that the mutual relationships between node pairs are correlated (or anti-correlated) whereas the uncorrelated relationships in the non-polarized state will result in an argument near zero.  This greater balance force for the polarized peace state drives its greater instability. We also note that the model dynamics enables a precise definition of the emergence of polarization via the appearance of a (polarized) first eigenvalue outside the noise band, which can be determined via random matrix theory. This enables prediction of how strong polarizing structure must be in order to affect stability.

That model-driven dynamics define the critical network structure for instability confers a level of generalizability difficult for qualitative theory, and metrics derived from it, to attain.  Note that it is the first eigenvalue that appears in the bifurcation condition, Eq.~(\ref{eq:alpha_PW_general}), underlying the polarization instability not a measure of polarization per se.  While not as immediately intuitive as polarization, eigenvectors and their eigenvalues are instrinsic properties of matrices, furnishing a natural coordinate basis for them, and hence networks as well. They also play a crucial role in dynamical systems theory \citep{strogatz_nonlinear_2016}. Consequently, this dynamics-defined critical structure allows for an integrated investigation of the impact upon stability of a broad array of different structural features, such as community structure, network size, degree distribution, and noise variance. It affords immediate prediction of the effects of such features via their impact on the first eigenvalue whereas it is often unclear how to infer their effects from qualitative theory. In addition, although qualitative reasoning can shed insight into how such features impact stability in isolation, it cannot readily integrate them theoretically or quantitatively. Furthermore, since the first eigenvalue is generated by the network itself, it can directly accommodate whatever cooperative and conflictual relationships are in the empirical network, without the need for a priori coarse graining as is often done in metrics customized for particular features; for example, Wayman's (\citeyear{Wayman1984}) bipolarization metric relies upon the counting of alliance cliques. Lastly, we point out that the ability to analytically derive the critical network structure dramatically facilitates the model's generalizability, obviating the need to laboriously conduct and synthesize simulation sweeps over multiple parameters.

The behavior of the first eigenvalue can be used to address the effect of increasing the network size on stability. We consider two separate effects, the behavior of the noise and that of the contrast signal producing polarized structure (Sec.~\ref{sec:commstruc}). If the network of dyadic biases is generated solely from noise with tie variance $\sigma^2$, the band edge is equal to $\sigma \sqrt{N}$ \citep{morrison_community_2020}. As the (positive) band edge in the noise only case is the leading eigenvalue, the critical balance sensitivity, $\alpha^*_{P \rightarrow W}$,  decreases as $1/\sqrt{N}$. Therefore, if we keep the variance constant, then noise-generated random structure will become less stable as the number of nodes is increased. This is similar to findings by May (\citeyear{may_will_1972}) that increasing connectivity in large complex systems decreases the system's stability. Allowing for signal, the contrast eigenvalue grows as $N$ so that polarized structure as well becomes less stable with increasing network size, as is also seen in the special case expression for $\alpha^*_{P \rightarrow W}$, Eq.~(\ref{eq:alpha_PW_special}). In addition, the faster growth of the contrast eigenvalue as compared with the noise implies that weaker polarizing processes (i.e., smaller $\mu$) can emerge as dominant structure for larger networks.

\subsection{Candidate variables for the bifurcation parameters} \label{sec:candidates}

We have shown how bifurcations can occur as our model parameters are changed. We now briefly discuss potential relationships between the parameters and variables considered in the international relations literature on systemic war. The dyadic bias parameters, representing longer time scale relationships of low-level cooperation and conflict between states are fairly straightforwardly associated with alliances and rivalries \citep{thompson_streetcar_2003}. The structural balance sensitivity, which determines how sharply the balance force rises before plateauing, is less clearly associated with a particular variable. In part, this ambiguity stems from the many variables scholars have put forward as underlying causes of systemic war. Any candidate for the balance sensitivity, however, must be associated with a dynamic whereby the decisions states make with respect to cooperation and conflict with other states becomes increasingly based on considerations of the extent to which they share friends and foes. We consider two such candidates: a shrinking power differential between the leading power and an ascending rival and the perceived dominance of offense over defense.

The decline in power of a dominant state relative to an ascending state has been claimed to be  a major source of systemic conflict \citep{KuglerOrganski1980, doran_war_1980, copeland_origins_2000}. The rising challenger's chafing at the status quo and the sense of threat felt by the leader create tension in the international system. Decreasing power differentials makes the dominant state less averse to war as its security is  threatened by changes to the status quo \citep{doran_war_1980}. The increasing specter of conflict generates pressure for other states to choose sides, thereby making structural balance dynamics more intense. This implies that the power differential between  the leading state and the challenger can potentially correspond to the structural balance sensitivity, with the peace-to-war bifurcation occurring near zero.

\begin{figure}[t]
    \centering
    \includegraphics[width=0.85\linewidth]{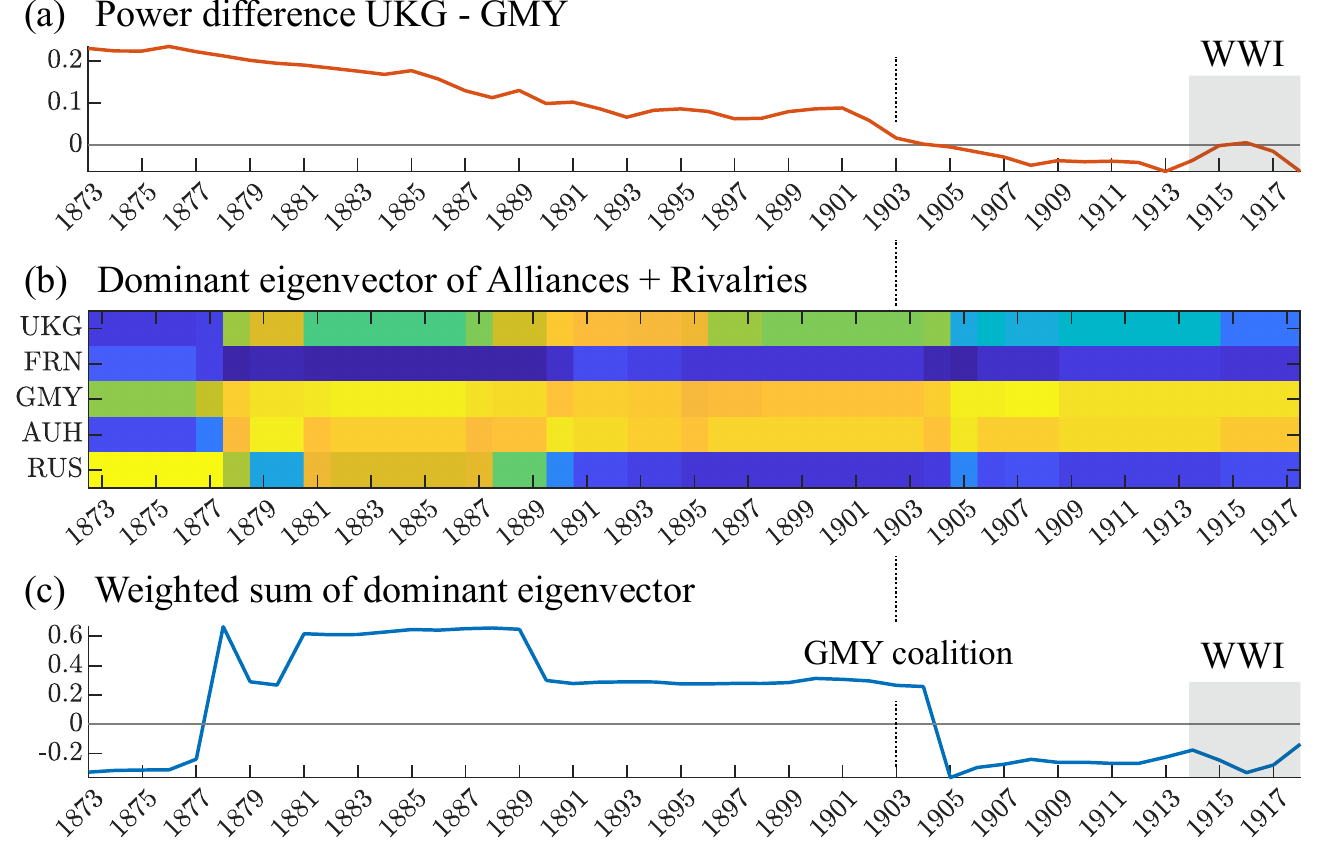}
    \caption{(a) Power difference between Great Britain and Germany. (b) Dominant eigenvector of the network of alliances and rivalries. (c) Sum of the dominant eigenvector weighted by relative power.}
    \label{fig:Euro_gps2}
\end{figure}

We illustrate the association between a narrowing power differential and factional alignments as represented in the first network eigenvector for the pre-WWI period in Fig.~\ref{fig:Euro_gps2}. The rise of Germany upon its consolidation in 1871 led to a steady shrinking of the power differential between it and Britain in the ensuing decades (Fig.~\ref{fig:Euro_gps2}(a)). The shifting factional alignments among the five great powers during this period can be observed in the leading eigenvector of the network  (Fig.~\ref{fig:Euro_gps2}(b)). The states with positive components (yellow shades) are aligned with Germany while those aligned against it have negative components (blue shades). Bismarck's skillful diplomacy in the years after consolidation prevented the rise of a counter-coalition. After his ouster in 1890, however, a clear anti-German alignment between Russia and France arose but Britain was not yet a member of this faction. This changed as Germany pursued a more provocative foreign policy along with a naval buildup that presented a threat to British maritime supremacy. In a manifestation of how intense structural balance pressures had become, to counter Germany, Britain abandoned its longtime policy of ``splendid isolation", agreeing to military cooperation with its colonialist rivals, France and Russia; the eigenvector components show Britain as belonging to the anti-German faction in 1905. Figure~\ref{fig:Euro_gps2}(c) shows how the power balance, as seen through the first eigenvector, between the pro and anti-German coalitions shifted over time. The component of the first eigenvector is weighted by the corresponding country's power and then the difference between the magnitudes of the sum of the pro-German (positive) components and the sum of the anti-German (negative) components is plotted. The balance tips in favor of the anti-German coalition in 1905, realizing German fears of encirclement and helping to set the stage for the war \citep{taylor_struggle_1954, Kissinger1994, kagan_origins_1995}.

Although the First World War did not start when Fig.~\ref{fig:Euro_gps2}(a) shows the power differential passing through zero in 1904, the war did start in 1914 when the differential was small, slightly in Germany's favor. In addition, the polarization of the system did increase in 1905 and some consider that the structural factors priming the system for war were already in place by then \citep{LevMul2021}. This is consistent with the fact that our model allows for war to be triggered in the bistable regime, prior to crossing the peace-to-war bifurcation threshold, as we discuss further below. Accordingly, the power differential can be viewed as a plausible driver for the balance sensitivity, particular given the considerable empirical support that power transition theory has received \citep{RaslerThompson2010}.

The perception among political and military leaders that offense dominates defense is another factor theorized as an important cause of systemic war \citep{VanEvera1999}. Paradoxically, although known for trench warfare and static frontlines, World War I is the paradigmatic example of this perception. Despite some recognition of how recent technological developments in firepower had advanced defense, an offensive doctrine in which superior elan and morale were the keys to victory dominated military thought among the great powers prior to the war \citep{Howard1986}. The belief that offense had the advantage fed the concern that, if a power left its ally unassisted when attacked by another great power, the ally would be quickly defeated, thereby leaving the power exposed to fight this belligerent alone \citep{christensen_chain_1990}. This worry encouraged ``chain ganging'' in which the great powers bound themselves more tightly to their allies and concomitantly assumed more confrontational stances toward their friends' foes. Conversely, a perceived advantage for defense loosens alliance commitments. Therefore, the extent to which warfare is viewed as favoring offense over defense, whether justified technologically or not, impacts structural balance pressures and so the offense-defense may be associated with the structural balance sensitivity. 

\subsection{Bifurcation structure and the catalysts of war} \label{sec:catalysts}

The bifurcation structure of our model can be used to more precisely formulate the debate about the nature of the triggers of systemic wars. The unstable quality of the international system prior to structural wars has been likened to a powder keg and spark, in which the former signifies the systemic conditions creating the instability, while the latter represents the immediate episode triggering the war. The debate, which has been most keenly engaged for World War I, concerns whether the spark must be drawn from a narrow set of specific catalysts or, instead, a wider, more generic set. In favor of a narrow set, Lebow (\citeyear{Lebow2000}) and Levy and Mulligan (\citeyear{LevMul2021}) argue that the onset of the First World War was highly contingent upon a crisis in the Balkans and, furthermore, upon the assassination of Archduke Franz Ferdinand, which removed the most influential voice of moderation in Austria-Hungary; if he had not been assassinated, the war might never have happened. On the other side, Thompson (\citeyear{thompson_streetcar_2003}) subscribes to the ``streetcar'' view of the war: the structural factors destabilizing Europe were so intense that, like missing a streetcar, even if one crisis passed without sparking a war, another one would have come along sooner or later to do the job. In essence, these two views differ as to whether the war can be considered contingent or inevitable.

The bistable dynamics exhibited in our model can accommodate both generic and specific paths to war. The generic route essentially corresponds to the system having reached the peace-to-war bifurcation in which the peace equilibrium vanishes, leaving only war (as occurs at $\alpha^*_{P \rightarrow W}$ in Fig.~\ref{fig:intro_fig}(c)). Here, any perturbation will cause the system to go from peace to war. Actually, it would be more precise to say that the system is just shy of the bifurcation so that a wide range of small (but not infinitesimal) perturbations will destabilize it. Turning to the specific route, that corresponds to the system being firmly in the bistable zone where both the peace and war equilibria are possible (say at $\alpha=0.03$ in Fig.~\ref{fig:intro_fig}(c)), the system requires a substantially larger kick to jump from peace to war. The system, however, may be particularly vulnerable to specific perturbation patterns; for example, the pattern that fully counterposes the Franco-Russian and Austrian-German alliances minimizes the required energy (Fig.~\ref{fig:ww1_control_sigs}(a)). If one sufficiently constrains the maximum size of a perturbation, then only a relatively narrow set of perturbations will shake the system into war. These most efficient perturbations then comprise the specific catalysts upon which war is contingent. 

Although the qualitative debate by necessity contrasts the war as either being inevitable or highly contingent, the model allows for a non-dichotomous formulation through use of the distance, in parameter space, from the peace-to-war bifurcation point. In the bistable zone, war onset is always considered to be contingent but the degree of contingency decreases as the system draws closer to the peace-to-war bifurcation point. In other words, the pool of destabilizing perturbations expands near the critical point. Note that, in the model, war is indeed inevitable beyond the peace-to-war bifurcation and, conversely, it is impossible prior to the war-to-peace bifurcation.

\subsection{Empirically testing war as bifurcation} \label{sec:testing}

The most fundamental prediction arising from the model is that the transition from peace to war, whether via the generic or specific triggers, involves bifurcation structure. Remarkably, this  prediction is amenable to empirical testing without the need to fit an explicit model to the data. This ability is a consequence of the universality of behavior for all systems in the vicinity of a given type of bifurcation, such as the saddle-node bifurcation displayed by our model. Critical transition theory identifies signatures associated with bifurcations, and critical slowing down in particular, such as increased variance, autocorrelations, and spatial correlations, longer recovery times, and power spectrum sharpening; the predictions involving not simply the direction of signature change, but their quantitative scaling as a function of distance from the bifurcation \citep{kuehn_mathematical_2011, scheffer_anticipating_2012, meisel_critical_2015, BurBauAna2020}.

Our preliminary look at the recovery rate response to militarized interstate disputes among the five great powers in the decades prior to World War I did not find evidence of critical slowing down. However, consistent with critical slowing down, Midlarsky (\citeyear{midlarsky_preventing_1984, midlarsky_hierarchical_1986}) theorized that as the onset of war grows closer, disputes would take longer to resolve and hence accumulate, finding evidence for such accumulation prior to World War I. Importantly, disputes involving lesser powers formed a critical part of his theory and findings. This suggests that future research on critical slowing down in international relations should involve wider networks beyond the great powers.

Given these mechanisms for destabilization, the formation of new alliances, increased connectivity, the addition of more countries to a system, or fluctuations in relative power could all increase the instability in a system of states and lead to systemic war.
While at the dyadic level, forming new connections may not add much conflict to a system or be detrimental for the parties involved, it could inadvertently make a system less stable by allowing for more nonlinear interactions between edges, increasing community structure, or equalizing a power imbalance.
Dynamical system models, such as the one introduced here, may be useful in assessing how changes to ties impact not only the countries involved but the entire system; models for system-wide impact could be beneficial in assessing systemic risk and designing de-escalation strategies.

\section{Conclusion} \label{sec:conclusion}

We have presented a model of the dynamics of cooperation and conflict that integrates complex systems and complex networks approaches. Our work makes contact with elements of complex systems theory including nonlinear dynamics, bifurcations, stability analysis, and critical transition signatures. With respect to complex networks, our work involves a signed networks, structural balance theory, and community structure. Spectral decomposition into eigenvalues and eigenvectors, and in particular the dimension reduction so afforded, provides a unifying framework for both the complex systems and networks aspects, allowing for calculation of bifurcation conditions, the compact representation of dominant polarized community structure, and empirical application to great power politics.

In our model, the tie value between two nodes evolves under the influence of a direct dyadic force biasing the nodes toward a low-level equilibrium, and a structural balance force, which shifts the nodes toward greater cooperation or conflict if they have similar or dissimilar relationships with mutual neighbors. We focused on the situation when the dyadic bias matrix allows for two possible stable equilibrium states: a peace state of low-level conflict and cooperation in which the equilibrium tie values are near their dyadic biases and a war state of high-level conflict and cooperation. Varying the structural balance sensitivity, which determines how sharply the balance force increases or decreases around the neutral point, then results in three regimes: a monostable peace state, a bistable regime where both peace and war states exist, and a monstable war state. Saddle-node bifurcations mark the transitions between these regimes. The sandwiching of the bistable regime between the peace and war monostable regimes results in a form of path dependence known as hysteresis: as the balance sensitivity is increased and then decreased past the peace-to-war bifurcation, the system does not return back down to the peace state at the same value it transitioned up to the war state, but rather a lower one.

We used our model to investigate structural features that make systemic war in a network more likely and identify patterns of perturbations that have the most impact on system stability.  We derived approximate analytical solutions for the bifurcation conditions in both a special case and generally, which reveal the key role of the first eigenvalue of the dyadic bias matrix. This enabled us to show how polarized community structure as contained in the first eigenvector makes the network more prone to systemic war by lowering the critical balance sensitivity at which the peace-to-war bifurcation occurs. This finding concords with theory emphasizing the destabilization effects of alliance polarization, but is also generalizable to structural features beyond polarization via the impact of those features on the first eigenvalue.

When the system is in the bistable regime, we considered the ability of finite, transient perturbations to induce a transition from peace to war. The most destabilizing perturbation pattern is constructed from the first eigenvector of the dyadic bias matrix. We also developed a scheme to find other destabilizing perturbations of low energy. We illustrated the empirical application of this perturbation analysis framework on the great power network immediately prior to World War I. The minimum energy perturbation excited cooperation and conflict, respectively, within and between the alliances involving the four Continental powers but not Great Britain. This pattern is consistent with the strong commitments exhibited within the Continental alliances in the crisis that triggered the war while Britain took a more ambiguous stance even though it was allied with France and Russia.

There are many empirical avenues for future research to build upon this work. The model's prediction of war as bifurcation could be tested via a systematic search for critical slowing down behaviors. The linkage between network structure and greater risk of systemic war could be found in network spectra as the emergence from the main spectral band of a polarized first eigenvector. Another line of empirical investigation would endeavor to identify variables in the international system that affect structural balance pressures, which could involve testing their correlation with a metric of structural balance \citep{kirkley_balance_2019, burghardt_dyadic_2020}. Although our substantive focus here has been international relations between states, other contexts involving cooperation and conflict driven by both dyad-specific and structural balance dynamics could serve as testing grounds, such as fragmented civil wars in which a multiplicity of armed groups fight and ally with each other.

We note a few ways in which the model can be extended. The products of the ties in the structural balance sum could be weighted by power so that powerful nodes contribute more to it than weaker ones. The ties could also be made directed to represent asymmetric relationships of aid or aggression. Other mechanisms influencing tie dynamics beyond the dyadic and structural balance forces could be incorporated into the model. One significant force that is not currently represented is power balancing dynamics in which state alliance decisions are driven by power calculations rather than existing alliances or enmities as in structural balance. States may ally with an erstwhile adversary to balance against a looming threat from an aggressive power. They may also abandon an existing ally to bandwagon with a powerful aggressor \citep{walt_origins_1990}. As maintaining the balance of power has long been claimed to be crucial to sustaining equilibrium in the international system, \citep{healy_balance_1973, Kissinger1994, RaslerThompson2010}, incorporating both power balancing and structural dynamics within the same model could be a fruitful way of better understanding how they interact to produce peace and war.

While this paper has concentrated on destabilizing factors and perturbations that make systemic war more likely, the model also has implications for preventing war and transitioning from war to peace. One example is that a perturbation pattern opposite to that calculated above as most dangerous will most efficiently keep the system from destabilizing. Another is that, in the bistable regime, optimal perturbations for kicking the system from the war to the peace state can be calculated rather than the reverse direction we treated herein. Alternatively, lowering the balance sensitivity beneath its critical value for the war-to-peace bifurcation would yield a monostable peace state. If the balance sensitivity were related to a power differential between key states, then increasing its war-to-peace critical value would correspond to lowering the power disparity at which peace is restored, implying that the losing side need not be defeated as decisively to bring the war to an end; Eq.~(\ref{eq:alpha_PW_general}) shows that decreasing the first eigenvalue of the dyadic bias matrix would have that effect. Finally, although it still lies in the realm of political philosophy and prophetic visions, at least for the international system, the model also possesses an equilibrium of high-level universal cooperation, which can be reached from both the peace and war states.


\appendix
\section{General treatment}
\label{app:general_case}

\renewcommand{\thefigure}{A\arabic{figure}}

\setcounter{figure}{0}

\numberwithin{equation}{section}
\setcounter{equation}{0}

We provide the derivations of the approximate bifurcation conditions stated in Sec.~\ref{sec:genresults} for a general dyadic bias matrix.

\subsection{Approximate system}

We replace the full matrix dynamical system, Eq.~(\ref{eq:dyn_sys_X_full_1}), with  analytically tractable forms by using approximations for the hyperbolic tangent function in the transition and plateau regions. In the transition region, $\theta=(\alpha / L) \b{X}^2 \ll 1$,  using $\tanh \theta \approx \theta$ yields
\begin{align} \label{eq:dyn_sys_X_approx}
    \frac{d\b{X}}{dt} = -\beta (\b{X} - \b{X}_D) +  \alpha \b{X}^2 + \b{X}_T(t).
\end{align}
This system has approximately the same stable peace and unstable fixed points as Eq.~\ref{eq:dyn_sys_X_full_1} for large $L$, but does not have the stable war fixed point.

In the plateau region, $|\theta| \gg 1$, we use $\tanh \theta \approx \pm 1$ to yield
\begin{align} \label{eq:dyn_sys_X_approx2a}
     \frac{d\b{X}}{dt} = -\beta (\b{X} - \b{X}_D) +  L\sign({\frac{\alpha}{L}\b{X}^2}) + \b{X}_T(t),
\end{align}
where the use of $\sign(\alpha{\b{X}^2}/L)$ assumes that all the components of $\b{X}^2$ are sufficiently large to be approximated as being on the positive or negative plateaus. 
We furthermore assume that $\sign({\b{X}^2})$ can be approximated using the signs of the components of the first eigenvector of $\b{X}$,  $\b{s}_1$. Denoting the vector of signs as $\b{u}$, we then have $\b{u} = \sign(\b{s}_1)$. We can then replace $\sign(\alpha{\b{X}^2/L})$ by $\b{u}\b{u}^T$ in Eq.~(\ref{eq:dyn_sys_X_approx2a}) yielding 
\begin{align} \label{eq:dyn_sys_X_approx2}
     \frac{d\b{X}}{dt} = -\beta (\b{X} - \b{X}_D) +  L\b{u}\b{u}^T + \b{X}_T(t).
\end{align}
This approximation is valid for $L$ sufficiently large so that, for trajectories heading from the peace state of low tie values to a high tie-value equilibrium, the feed-forward growth has time to preferentially amplify the leading eigenvector $\b{s}_1$, similar to the behavior in the pure structural balance model of Eq.~\ref{eq:marvel}.


\subsection{Stable and unstable fixed points}


We now find fixed points of the approximate system in the transition region, Eq.~(\ref{eq:dyn_sys_X_approx}), in the absence of the transient term ($\b{X}_T=0$) by expressing it in terms of the dynamics of its eigenvalues and eigenvectors. We denote the tie matrix at a fixed point by $\b{X}_0$ and write its eigenvector decomposition  as $\b{X}_0 = \b{S}_0 \Lambda_0 \b{S}_0^T$ and that of the dyadic bias matrix as $\b{X}_D = \b{S}_D \Lambda_D \b{S}_D^T$. For simplicity, we assume that $\b{X}_0$ and $\b{X}_D$, are full-rank matrices, allowing us to perform an eigendecomposition. Setting $\frac{d\b{X}}{dt} = 0$, using these decompositions, and rearranging yields
\begin{align}
    0 &= -\beta \left(\b{X}_0 - \b{X}_D \right) + \alpha \b{X}_0^2\\
    0 &= \alpha \b{X}_0^2 -\beta \b{X}_0 + \beta \b{X}_D \\
    0 &= \alpha \b{S}_0 \Lambda_0^2 \b{S}_0^T -\beta \b{S}_0 \Lambda_0 \b{S}_0^T + \beta \b{S}_D \Lambda_D \b{S}_D^T\\
      \b{S}_0(\alpha \Lambda_0^2 -\beta \Lambda_0) \b{S}_0^T &= - \beta \b{S}_D \Lambda_D \b{S}_D^T,
\end{align}
where we have used the orthonormality of the eigenvectors, $\b{S}_0^T \b{S}_0 = I$ in the third line.

We diagonalize the left hand side by multiplying by $\b{S}_0^T$ and $\b{S}_0$ on the left and right respectively,
\begin{align}
     \alpha \Lambda_0^2 -\beta \Lambda_0 &= - \beta \b{S}_0^T\b{S}_D \Lambda_D \b{S}_D^T \b{S}_0\\
     \alpha \Lambda_0^2 -\beta \Lambda_0 &= - \beta \b{D} \Lambda_D \b{D}^T,
     \label{eq:lambda_pre}
\end{align}
where $\b{D} = \b{S}_0^T\b{S}_D$. Since the left hand side is diagonal then the right hand side must also be diagonal. Therefore $\b{D}$ must be the identity matrix, $\b{S}_0^T\b{S}_D = I$. Therefore $\b{S}_0 = \b{S}_D$.
This implies that the eigenvectors of $\b{X}_0$ are the same as the eigenvectors of $\b{X}_D$. This allows us to diagonalize the matrix dynamical system equation at the fixed points and solve for the eigenvalues of $\b{X}$ there. As the righthand side of (\ref{eq:lambda_pre}) simplifies to $-\beta \Lambda_D$, we rearrange to obtain 
\begin{align}
    0 &= \alpha \Lambda_0^2 -\beta \Lambda_0 + \beta \Lambda_D,
\end{align}
which expressed in terms of the individual eigenvalues $\lambda_{0i}$ is
\begin{align}
    0 &= \alpha \lambda_{0i}^2 -\beta \lambda_{0i} + \beta \lambda_{Di}.
\end{align}
Solving yields
\begin{align}
    \lambda_{0i} &= \frac{\beta \pm \sqrt{\beta^2 - 4\alpha \beta \lambda}_{Di}}{2 \alpha}  ,\quad i = 1,...,N \label{eq:eigval_fps}
\end{align}
for the fixed point eigenvalues. Eigenvalues of the stable fixed point are given by the negative sign and those of the unstable fixed point by the positive sign as we now show.

For a stable fixed point, small perturbations will decay whereas they will grow for an unstable fixed point. The response is given by the eigenvalues of the linearized system. Denoting the perturbation by $\b{\delta X}$, the linearized system is
\begin{align} \label{eq:linear}
    \frac{d \b{\delta X}}{dt} &= -\beta \b{\delta X} + \alpha (\b{X}_0 \b{\delta X} + \b{\delta X} \b{X}_0).
\end{align}
Using the above result that the eigenvectors of $\b{X}_0$ are the same as $\b{X}_D$, we expand the fixed point and the perturbation in terms of the eigenvectors, $\b{s}_D$, so that $\b{X}_0 = \sum_{i=1}^N \lambda_{0i} \b{s}_{Di}\b{s}_{Di}^T$ and $\b{\delta X}_0 = \sum_{i=1}^N a_i(t) \b{s}_{Di}\b{s}_{Di}^T$, where $a_i(t)$ is the time-dependent amplitude of the $i^{th}$ mode. Using these expansions in Eq.~(\ref{eq:linear}), we find that each mode evolves independently via
\begin{align} \label{eq:modede}
    \frac{d a_i}{dt} &= \left(-\beta + 2\alpha \lambda_{0i}\right) a_i.
\end{align}
Taking the modes to have an exponential time dependence, $a_i \sim e^{\nu_i t}$, the growth rate $\nu_i$ is given by
\begin{align} \label{eq:growthrate}
    \nu_i &= -\beta + 2\alpha \lambda_{0i}.
\end{align}
Substituting for $\lambda_{0i}$ using (\ref{eq:eigval_fps}) then gives
\begin{align}
    \nu_i &= \pm \sqrt{\beta^2 - 4\alpha \beta \lambda}_{Di}. 
\end{align}
The positive and negative signs in Eq.~(\ref{eq:eigval_fps}) therefore correspond, respectively, to growing and decaying perturbations and so are associated with the unstable and stable fixed points as claimed above.

The stable fixed point corresponds to the peace state and its leading eigenvalue $\lambda_P$ is given by
\begin{align}\label{eq:lambda_P_notrans}
    \lambda_P = \frac{\beta - \sqrt{\beta^2 - 4\alpha \beta \lambda}_{D1}}{2 \alpha}.
\end{align}
For the unstable fixed point, the leading eigenvalue is not of concern but rather the lowest one,
\begin{align} 
    \lambda_U &= \frac{\beta + \sqrt{\beta^2 - 4\alpha \beta \lambda}_{D1}}{2 \alpha},
    \label{eq:lambda_U_notrans}
\end{align}
as the peace-to-war bifurcation occurs when it equals $\lambda_P$.

The case when a finite, transient perturbation, $\b{X}_T(t)$, is applied to the system can be treated similarly if we take it to have the form of an impulse where all its elements are constant over the duration of the transient. Defining the modified dyadic bias matrix, $\tilde{\b{X}}_D(t) = \b{X}_D + \b{X}_T(t)/\beta$, we can rewrite Eq.~(\ref{eq:dyn_sys_X_approx}) as
\begin{align}
    \frac{d\b{X}}{dt} &= -\beta \left(\b{X} - \tilde{\b{X}}_D(t) \right) + \alpha \b{X}^2.
\end{align}
This modification in dyadic biases results in a new effective peace state eigenvalue given by
\begin{align}\label{eq:lambda_P_trans}
    \lambda_P = \frac{\beta - \sqrt{\beta^2 - 4\alpha \beta \tilde{\lambda}_{D1}(t)}}{2 \alpha}.
\end{align}
Similarly, the lowest eigenvalue of the unstable state becomes
\begin{align}
    \lambda_U &= \frac{\beta + \sqrt{\beta^2 - 4\alpha \beta \tilde{\lambda}_{D1}(t)}}{2 \alpha}.
    \label{eq:lambda_U_trans}
\end{align}
If the perturbation is sufficiently small, then $\lambda_P$ and $\lambda_U$ will be real and nonzero. The corresponding fixed points will exist and, in particular, the system will settle into a new peace state as shifted by the perturbation, assuming that the transient duration is long enough. A large perturbation, however, can cause the square root in (\ref{eq:lambda_P_trans}) and (\ref{eq:lambda_U_trans}) to be imaginary, in which case the stable and unstable fixed points do not exist; the system will head toward a stable equilibrium of high conflict and/or cooperation as we now discuss.

\begin{figure}[t]
    \centering
    \includegraphics[width=0.8\linewidth]{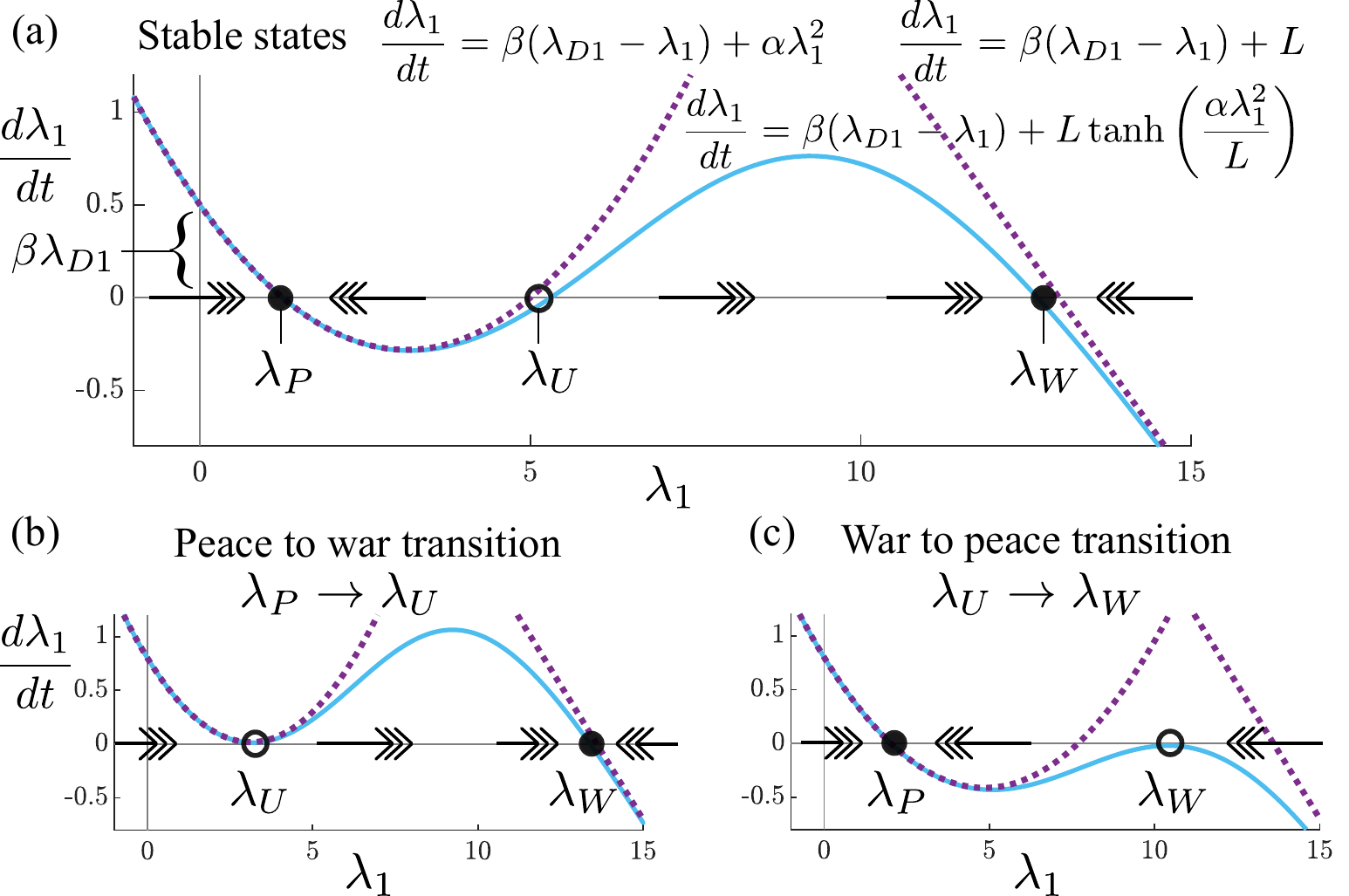}
    \caption{(a) Dynamics of the leading eigenvalue as a function of initial conditions. $\lambda_P$ and $\lambda_W$ are stable fixed points separated by an unstable fixed point $\lambda _U$. Curves approximate the eigenvalue dynamics. $\alpha=0.08$, $\beta = 0.5$, $L = 6$, and $\lambda_{D1} = 1$. (b) A saddle-node bifurcation of the stable and unstable fixed points $\lambda_P$ and $\lambda_U$ underlies the peace to war transition. $\alpha = 0.08$, $\beta = 0.5$, $L=6$, and $\lambda_{D1}=1.6$.
      (c) A saddle-node bifurcation of $\lambda_W$ and $\lambda_U$ underlies the war to peace transition. $\alpha=0.052$, $\beta=0.5$, $L=6$, and $\lambda_{D1}=1.6$. Arrows shows the direction of movement of the eigenvalues.
      }
    \label{fig:hysteresis}
\end{figure}


We calculate the high conflict and cooperation fixed points using the plateau approximation of the dynamics as given by Eq.~(\ref{eq:dyn_sys_X_approx2}). Setting $\frac{d\b{X}}{dt}=0$ and solving for the fixed point $\b{X}_0$ produces
\begin{align}
    \b{X}_0 &= \b{X}_D + \frac{L \b{u}\b{u}^T + \b{X}_T(t)}{\beta}.
\end{align}
When no transient perturbation is applied, the stable state which results from the peace-to-war bifurcation will be the war state as determined by the leading eigenvector of $\b{X}_D$ and we write the sign vector as $\b{u}_{D1}=\sign(\b{s}_{D1})$. The war state in the plateau approximation is then given by
\begin{align} \label{eq:war_state}
        \b{X}_W &= \b{X}_D + \frac{L}{\beta} \b{u}_{D1}\b{u}_{D1}^T
\end{align}

When an applied perturbation does shift the system out of the peace state, the first eigenvector, $\tilde{\b{s}}_{D1}$, of the modified dyadic bias matrix will grow to dominate the network if $L/\beta$ is large in comparison with $||\tilde{\b{X}}_D||$. In this case, after the perturbation has ended, the final high tie value state of the system, $\b{X}_H$, will be approximately
\begin{align} \label{eq:high_state}
        \b{X}_H &= \frac{L}{\beta} \tilde{\b{u}}_{D1}\tilde{\b{u}}_{D1}^T,
\end{align}
where $\tilde{\b{u}}_{D1}=\sign(\tilde{\b{s}}_{D1})$. This expression was used to calculate the final states in Fig.~\ref{fig:ww1_control_sigs}. However, if $L$ is not large enough to provide enough time for  $\tilde{\b{u}}_{D1}$ to grow to dominate the network, the final network structure will have noticeable contributions from other eigenvectors of $\tilde{\b{X}}_D$.

\subsection{Bifurcation transition from peace to war}

The peace-to-war transition occurs when the first eigenvalue of the linearized system at the peace state is zero, corresponding to a saddle-node bifurcation. Setting $\nu_1=-\sqrt{\beta^2 - 4\alpha \beta \lambda}_{D1}$ to zero and solving for the critical balance sensitivity yields
\begin{align} 
    \alpha^*_{P\rightarrow W} = \frac{\beta}{4 \lambda}_{D1}. \label{eq:alpha_star}
\end{align}
The vanishing of the discriminant in Eq.~(\ref{eq:lambda_P_notrans}) yields the critical value of the peace state eigenvalue at the bifurcation,
\begin{align} \label{eq:lambda_P*}
    \lambda_{P}^* = \frac{\beta}{2 \alpha^*_{P \rightarrow W}}.
\end{align}
Equation~(\ref{eq:lambda_U_notrans}) shows that the unstable state eigenvalue also has this value at the bifurcation so that $\lambda_P^* = \lambda_U^*$. This equality of the stable and unstable critical eigenvalues corresponds to the merging of the stable and unstable fixed points that occurs in a saddle-node bifurcation.

For the case where the impulse perturbation $\b{X}_T$ is applied to the peace equilibrium $\b{X}_P$, the system parameters $\alpha$ and $\beta$ are fixed and we seek the critical eigenvalue, $\tilde{\lambda}_{D1}^*$, of the modified dyadic bias matrix $\tilde{\b{X}}_D$. This will occur when the effective peace and unstable eigenvalues due to the perturbation are equal. This yields a condition analogous to (\ref{eq:alpha_star}) but which we express in terms of the critical eigenvalue, 
\begin{align}\label{eq:lam_tilde_star}
     \tilde{\lambda}_{D1}^* &= \frac{\beta}{4 \alpha}.
\end{align}
If $\tilde{\lambda_{D1}} > \tilde{\lambda}_{D1}^*$, then the perturbation will kick the system out of the peace state into a high conflict and cooperation state.

\subsection{Bifurcation transition from war to peace}

We consider the bifurcation from the war state, $\b{X}_W$, back to the peace state as illustrated in the hysteresis curve of Fig.~\ref{fig:intro_fig}(c). No external perturbation is imposed in this case ($\b{X}_T=0$). Equation~(\ref{eq:war_state}) shows that the war state is the sum of $\b{X}_D$ and a rank-one matrix. We therefore approximate $\lambda_W$ by the expression for the upper bound of the leading eigenvalue of the sum of a symmetric matrix and a rank-one modification \citep{bunch_rank-one_1978},
\begin{align}
    \lambda_W = \lambda_{D1} + \frac{L N}{\beta}.
    \label{eq:lambda_war}
\end{align}
Note that increasing the size of the network as well as $L/\beta$ increases $\lambda_W$. A larger $\lambda_{D1}$ has similar effect implying that a polarized peace state is not only more susceptible to war (Sec.~\ref{sec:commstruc}) but results in more intense conflict should that war arise.  

The transition from the war to the peace state is also generated by a saddle-node bifurcation, one distinct from the peace-to-war transition. The war-to-peace bifurcation occurs when the stable war fixed point and the unstable fixed point collide, leaving only the stable peace state as the remaining equilibrium. Given that our approximate dynamical systems in the transition and plateau regions do not mesh smoothly, we cannot perform a linear stability analysis for the war state (which is always stable in the plateau approximation) as we did for the peace state. However, we can make a rough calculation of the bifurcation condition by setting $\lambda_W$ equal to $\lambda_U$ from Eq.~(\ref{eq:lambda_U_notrans}). Doing so and solving for the critical balance sensitivity yields
\begin{align}
    \lambda_{D1} + \frac{L N}{\beta} &= \frac{\beta + \sqrt{\beta^2 - 4\alpha \beta \lambda_{D1}}}{2 \alpha}\\
    	\alpha^*_{W\rightarrow P} &= \frac{NL}{(\lambda_{D1} + NL/\beta)^2}.
\end{align}
Thus, for $L/\beta$ large, the transition back from war to peace will take place at a much lower $\alpha$ value than for the transition from peace to war, Eq.~(\ref{eq:alpha_star}).

Figure~\ref{fig:hysteresis} visualizes the bifurcation structure of our system as approximated by the dynamics of the first eigenvalue, $\lambda_1(t)$, under the expansion of the network by the dyadic bias matrix eigenvectors, $\b{X}(t)=\sum_k \lambda_i(t) \b{s}_{Dk} \b{s}_{Dk}^T$. Its rate of change, $d\lambda_1/dt$, is plotted as a function of $\lambda_1$ for the transition and plateau approximations (dotted curves) for which the modes evolve independently. Also shown is as a smooth approximation that retains the hyperbolic tangent (solid curve). Points where $d\lambda_1/dt=0$ correspond to fixed points and the arrows indicate their stability. Figure~\ref{fig:hysteresis}(a) shows the bistable regime in which the $\lambda_1$ is stable at $\lambda_P$ and at $\lambda_W$, with the unstable fixed point residing between them.   For an increased value of $\lambda_{D1}$,  Figure~\ref{fig:hysteresis}(b) shows the saddle-node bifurcation point corresponding to the peace-to-war transition in which $\lambda_P$ and $\lambda_U$ have merged. The other saddle-node bifurcation, in which $\lambda_U$ merges with $\lambda_W$, is shown in Fig.~\ref{fig:hysteresis}(c) and corresponds to the war-to-peace transition when $\alpha$ is decreased.

\section*{Acknowledgement}
This work was supported by the National Science Foundation (MM, award no. 2103239); the Army Research Office (MM and MG, grant no. W911NF1910291); and the Air Force Office of Scientific Research (JNK, grant no. FA9550-
17-1-0329).



\end{document}